\documentclass[
 reprint,
superscriptaddress,
 amsmath,amssymb,
 aps,
prx,
]{revtex4-2}
\usepackage{graphicx}
\usepackage{dcolumn}
\usepackage{bm}
\usepackage[mathlines]{lineno}%
\usepackage[utf8]{inputenc}
\usepackage[T1]{fontenc}
\usepackage{mathptmx}
\usepackage{amsmath}
\usepackage{etoolbox}
\usepackage{float}
\usepackage{nicefrac}
\usepackage[normalem]{ulem}

\usepackage{siunitx}
\usepackage{xcolor}
\usepackage{esint}
\usepackage{booktabs}

\DeclareSymbolFont{newfont}{OML}{cmm}{m}{it}

\DeclareMathSymbol{\Epsilon}{0}{newfont}{15}
\DeclareMathSymbol{\Rho}{0}{newfont}{37}

\renewcommand\epsilon{\Epsilon}
\renewcommand\varrho{\Rho}

\usepackage{xr}

\newcommand{\hypgeo}[2]{%
  {\vphantom{F}}_{#1}\kern-\scriptspace F_{#2}%
}
\def\onedot{$\mathsurround0pt\ldotp$}
\def\cdddot#1{
  \mathbin{\vcenter{\baselineskip.67ex
    \hbox{\onedot}\hbox{\onedot}\hbox{\onedot}%
  }}%
}
\definecolor{changed}{rgb}{0,0,0}
\definecolor{lightGrey}{rgb}{0.6,0.6,0.6}
\definecolor{darkGrey}{rgb}{0.3,0.3,0.3}
\definecolor{codegreen}{rgb}{0,0.6,0}
\definecolor{codegray}{rgb}{0.5,0.5,0.5}
\definecolor{codepurple}{rgb}{0.58,0,0.82}
\definecolor{backcolour}{rgb}{0,0,0}
\definecolor{mBlue}{rgb}{0.12, 0.47, 0.71}
\definecolor{mRed}{rgb}{0.69, 0.13, 0.13}

\makeatletter
\def\@email#1#2{%
 \endgroup
 \patchcmd{\titleblock@produce}
  {\frontmatter@RRAPformat}
  {\frontmatter@RRAPformat{\produce@RRAP{*#1\href{mailto:#2}{#2}}}\frontmatter@RRAPformat}
  {}{}
}%
\makeatother
\begin{document}
\title[]{Complementary Eigen–Zundel Interpretation Reconciles Thermodynamics and Spectroscopy of Excess Protons in Aqueous HF Solutions}

\author{Louis Lehmann}
\affiliation{Department of Physics, Freie Universität Berlin, Arnimallee 14, 14195 Berlin, Germany}

\author{Florian N. Brünig}
\affiliation{Department of Physics and Materials Science, University of Luxembourg, L-1511 Luxembourg City, Luxembourg}

\author{Jonathan Scherlitzki}
\affiliation{Department of Physical and Theoretical Chemistry, Freie Universität Berlin, Arnimallee 22, 14195 Berlin, Germany}

\author{Morten Lehmann}
\affiliation{Institute of Chemistry, Technische Universität Berlin, Straße des 17. Juni 115, 10623 Berlin, Germany}

\author{Martin Kaupp}
\affiliation{Institute of Chemistry, Technische Universität Berlin, Straße des 17. Juni 115, 10623 Berlin, Germany}

\author{Beate Paulus}
\affiliation{Department of Physical and Theoretical Chemistry, Freie Universität Berlin, Arnimallee 22, 14195 Berlin, Germany}

\author{Roland R. Netz$^*$}
\email{corresponding author: rnetz@physik.fu-berlin.de}
\affiliation{Department of Physics, Freie Universität Berlin, Arnimallee 14, 14195 Berlin, Germany}

\begin{abstract}
 Aqueous solutions of HF and HCl behave very differently at intermediate concentrations: HCl dissociates completely, whereas HF remains only partially dissociated and forms bifluoride (HF$_2^-$). 
This should lead to markedly different excess-proton spectra in HF and HCl solutions, in contrast to experimental reports.
Using ab-initio molecular dynamics, we show that in HF the proton is not firmly bound to F$^-$, as suggested by textbook chemistry, but dynamically shared with a hydrating water molecule. This is rationalized by a modified Eigen-picture description which also explains the formation of HF$_2^-$. The similar vibrational spectra of HF and HCl solutions are explained by a complementary Zundel picture in terms of almost identical excess proton transfer free-energy profiles for HF and HCl.
These results reconcile thermodynamic and spectroscopic observations and provide a unified microscopic picture of excess protons in aqueous solution.
\end{abstract}
\maketitle
Understanding the behavior of excess protons in aqueous solution is of fundamental importance in many areas of chemistry and biology, such as
proton-coupled electron transfer processes \cite{warburtonTheoreticalModelingElectrochemical2022,basergaProtonCoupledElectronTransfer2025},
electrocatalytic reactions \cite{liMechanismCationsSuppressing2023},
enzymatic reactions \cite{siegbahnHowProtonsMove2023},
membrane proton diffusion \cite{heberleProtonMigrationMembrane1994,agmonProtonsHydroxideIons2016a},
and fuel cells \cite{weeApplicationsProtonExchange2007,jiaoDesigningNextGeneration2021}. 
The thermodynamic properties of an acid in aqueous solution are well described by the acid dissociation reaction
\begin{equation}
\mathrm{HA}\rightleftharpoons \mathrm{A^-} + \mathrm{H^+}\, ,
\label{eq:Bronsted}
\end{equation}
where HA is the acid molecule, A$^-$ its conjugated base and H$^+$ is the excess proton \cite{atkinsAtkinsPhysicalChemistry}. Strong acids such as HCl ($\mathrm{A}^-=\mathrm{Cl}^-$, $\mathrm{p}K_a=-7.0$) dissociate almost completely in water, while weak acids such as HF ($\mathrm{A}^-=\mathrm{F}^-$, $\mathrm{p}K_a=3.2$) only partially \cite{atkinsAtkinsPhysicalChemistry}.  
Measurements of electromotive force \cite{broeneThermodynamicsAqueousHydrofluoric1947,braddyEquilibriaModeratelyConcentrated1994}, conductivity \cite{daviesXXXVTransferenceNumbersIonic1924,hamerActivityCoefficientsHydrofluoric1970}, the Hammett acidity function \cite{bell262KineticsDepolymerisation1956,hymanHammettAcidityFunction1957}, and calorimetry \cite{heplerHeatEntropyIonization1953} indicate that at moderate HF concentrations F$^-$ is converted into bifluoride (HF$_2^-$) according to
\begin{align}
  \mathrm{HF} + \mathrm{F}^{-} \rightleftharpoons \mathrm{HF_2^-},
  \label{eq:bifluoride}
\end{align}
which increases the activity of H$^+$ according to Le Chatelier's principle.
It has been known for almost a century that acids with vastly different dissociation behavior exhibit similar IR spectra \cite{plylerInfraredAbsorptionAcid1934a}. Solutions of HF and HCl in particular where investigated by Giguère \textit{et al.} \cite{giguereH3OIonsAqueous1976,giguereIonPairsStrength1976,giguereNatureHydrofluoricAcid1980} who found no spectral evidence for HF$_2^-$.
Instead, they proposed that the observed spectra can be explained by a Zundel-like H$_3$OF configuration in which the proton is shared between water and fluoride. 
\\ \\
Previous ab-initio molecular dynamics (AIMD) simulations of HF solutions \cite{laasonenInitioMolecularDynamics1996,sillanpaaStructuralSpectralProperties2002,simonInitioMolecularDynamics2005,iftimieSpectralSignaturesMolecular2008,thomasUnderstandingDissociationWeak2009} identified a significant population of H$_3$OF configurations, while no formation of HF$_2^-$ ions was observed at moderate concentrations, consistent with Giguère's interpretation of experimental spectra. Simon and Klein \cite{simonInitioMolecularDynamics2005} observed spontaneous HF$_2^-$ formation in AIMD simulations of an equimolar HF–water mixture, demonstrating that HF$_2^-$ formation is accessible on AIMD timescales, but not favored at moderate concentrations, in apparent contradiction with thermodynamic predictions.
Rather than invoking HF$_2^-$ formation, Sillanpää et al. \cite{sillanpaaStructuralSpectralProperties2002} proposed that the increasing acidity at higher concentrations can be partly attributed to enhanced fluoride stabilization arising from the increasing ionic strength of the solution. This interpretation, however, is not fully consistent with Debye–Hückel theory, which predicts that increased ionic strength stabilizes all charged species and thus lowers the activity of excess protons as well.
\\ \\
In strong acids, excess-proton hydrates are typically interpreted as Zundel (H$_5$O$_2^+$) or Eigen (H$_9$O$_4^+$) cations, representing limiting cases of proton sharing between two water molecules or proton localization on a hydronium ion stabilized by three hydrogen-bonded water molecules \cite{marxNatureHydratedExcess1999,schmittComputerSimulationProton1999,dahmsLargeamplitudeTransferMotion2017,fournierBroadband2DIR2018,kunduHydratedExcessProtons2019,carpenterDecoding2DIR2020,calioMolecularOriginsBarriers2020,calioResolvingStructuralDebate2021,schroderCouplingHydratedProton2022,gomezNeuralnetworkbasedMolecularDynamics2024,dipinoZundEigStructureProton2023,yangUnveilingIntermediateHydrated2025,maRevisitingH5O2IR2025}. 
The minimal motif that explains spectral signatures of excess protons in strong acids is Zundel-like, with an excess proton shared between two water molecules\cite{brunigSpectralSignaturesExcessproton2022,schroderCouplingHydratedProton2022}, which was utilized by Brünig et al. to identify spectral signatures of proton transfer \cite{brunigSpectralSignaturesExcessproton2022}.
However, structurally excess proton hydrates in strong acids more closely resemble distorted Eigen-cations, whereas Zundel-like configurations are commonly interpeted as transition states during proton transfer \cite{tuckermanInitioMolecularDynamics1995,schmittComputerSimulationProton1999,markovitchSpecialPairDance2008,berkelbachConcertedHydrogenBondDynamics2009,calioResolvingStructuralDebate2021}. At higher HCl concentrations, chloride can replace a water molecule in the distorted Eigen cation \cite{royResolvingHeterogeneousDynamics2020,calioMolecularOriginsBarriers2020}, creating an entropic barrier for proton transport \cite{calioMolecularOriginsBarriers2020}.
Excess protons have been experimentally extensively characterized by infrared (IR) spectroscopy \cite{giguereIonPairsStrength1976,khoramiEtudeMelangesEau1987,kimVibrationalSpectrumHydrated2002,leveringObservationHydroniumIons2007,thamerUltrafast2DIR2015,
  dahmsLargeamplitudeTransferMotion2017,
  kunduHydratedExcessProtons2019,
  carpenterPicosecondProtonTransfer2018,zengDemystifyingDiffuseVibrational2021,ekimovaLocalCovalentBonding2022,lounasvuoriVibrationalSignatureHydrated2023}.
Multistate empirical valence bond simulations \cite{schmittComputerSimulationProton1999,kimVibrationalSpectrumHydrated2002,calioResolvingStructuralDebate2021}, AIMD \cite{iftimieMolecularOriginContinuous2006,napoliDecodingSpectroscopicFeatures2018,brunigSpectralSignaturesExcessproton2022}, and machine-learning molecular dynamics \cite{beckmannInfraredSpectraCoupled2022,atsangoDevelopingMachinelearnedPotentials2023,gomezNeuralnetworkbasedMolecularDynamics2024,yangUnveilingIntermediateHydrated2025} have reproduced characteristic excess-proton IR signatures in strong acids, in some cases even without including the conjugate base \cite{beckmannInfraredSpectraCoupled2022,atsangoDevelopingMachinelearnedPotentials2023,gomezNeuralnetworkbasedMolecularDynamics2024}. AIMD studies of HF report substantially different IR broad-band signatures for chemical species 
appearing in HF solutions 
\cite{sillanpaaStructuralSpectralProperties2002,
iftimieMolecularOriginContinuous2006}.
Other studies focused on cryogenic temperatures and found pronounced spectral differences between the hydrohalic acids and proton-sharing intermediates \cite{iftimieSpectralSignaturesMolecular2008,thomasUnderstandingDissociationWeak2009}. 
\\ \\
Thermodynamics predicts markedly different chemical speciation in aqueous HF and HCl, whereas their experimental spectra are very similar, which is puzzling.  We analyze AIMD simulations by an extended Eigen framework, which yields chemical speciation consistent with thermodynamic expectations, including bifluoride formation.
Spectral proton transfer signatures are best understood in a complementary Zundel picture, by which we explain why the proton-transfer bands of HF and HCl solutions are very similar under ambient conditions and typical experimental concentrations (7.5~mol/kg), even though the majority of excess protons in HF are in direct contact with fluoride, while in HCl they are mainly shared between water molecules. The puzzling similarity arises because electrostatic screening renders the free-energy landscapes within the Zundel configurations H$_5$O$_2^+$ (dominant in HCl) and H$_3$OF (dominant in HF) nearly indistinguishable, leading to similar proton-transfer dynamics and thereby similar IR signatures of proton transfer. Throughout, we use the Eigen and Zundel pictures as complementary views of the excess proton: the Eigen picture describes a central hydronium ion hydrogen-bonded to three proton acceptors, whereas the Zundel picture focuses on proton sharing of the hydronium's most weakly bound proton with a single proton acceptor. 
\section*{Results and Discussion}
\label{results_discussion}
\begin{figure*}
    \centering
     \includegraphics[width=1.\textwidth]{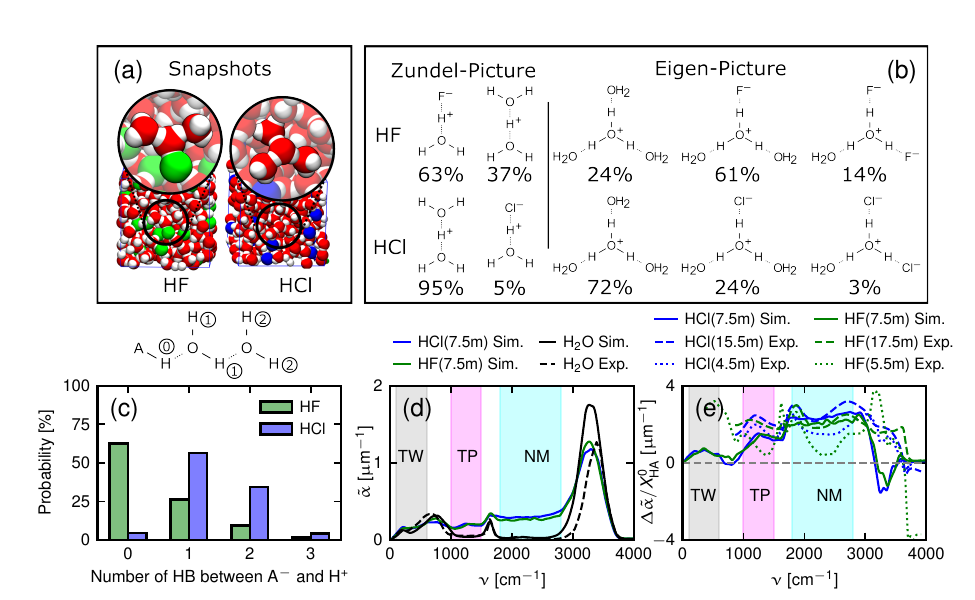}
    \caption{
      Analysis of excess-proton solvation, and IR spectra of 7.5~mol/kg HF and HCl solutions.
       (a) Simulation snapshots of systems containing 224 water molecules and 30 acid molecules, with HF shown on the left and HCl on the right, each including a zoomed view on representative Eigen structures.
  (b) Chemical structures and probability distribution of excess proton solvation in the Zundel-picture and Eigen-picture for the two acid solutions. 
   (c) Probability of finding an excess proton (H$^+$) and its conjugate base (A$^-$) in HF and HCl separated by a minimal number of hydrogen bonds, as indicated by the circled numbers in the sketch above the panel.
       (d) Simulated absorption coefficients $\tilde \alpha(\omega)$ of pure water, and HF and HCl solutions.
      The experimental absorption coefficient of pure water \cite{haleOpticalConstantsWater1973b} is shown for comparison.
      (e) Rescaled difference spectra, as defined in Eq.~\eqref{eq:difference_absorb}, for the simulated acid solutions compared with experimental difference spectra extracted from the literature: HF (5.5~m) \cite{khoramiEtudeMelangesEau1987}, HCl (4.5~m) \cite{thamerUltrafast2DIR2015}, HCl (15.5~m)\cite{giguereH3OIonsAqueous1976}, HF (17.5~m)\cite{giguereH3OIonsAqueous1976}.
      The colored frequency ranges correspond to the transfer-waiting (TW, gray), transfer-path (TP, magenta), and normal-mode (NM, cyan) regions \cite{brunigSpectralSignaturesExcessproton2022}. Concentrations are given in units of molality m$=$mol/kg.
      }
    \label{fig:HF_HCl_Spectra}
  \end{figure*}
  \subsection*{Infrared Spectra and Chemical Speciation}
  We begin our analysis by comparing the solvation structure of excess protons in 7.5~mol/kg hydrofluoric acid (HF) and hydrochloric acid (HCl) solutions at ambient conditions using AIMD simulations. Simulation snapshots are shown in Fig. \ref{fig:HF_HCl_Spectra}(a), along with a zoom into the prevalent excess-proton hydration structures, which are mainly coordinated by fluoride and water in HF solutions, and mainly by water in HCl solutions. 
  \\ \\
  An analysis of proton solvation beyond the Zundel–Eigen picture is shown in Fig. \ref{fig:HF_HCl_Spectra}(c), which reports the probability of finding the excess proton and its conjugate base separated by the minimal number of hydrogen bonds (i.e., along the shortest hydrogen-bond path), as illustrated schematically above the panel.
  In HF, fluoride most frequently remains in direct contact with the excess proton (0 hydrogen bonds). Each additional hydrogen bond roughly halves this probability, and separations beyond two hydrogen bonds are rare, limiting the participation of excess protons in electrode reactions 
  \cite{broeneThermodynamicsAqueousHydrofluoric1947,braddyEquilibriaModeratelyConcentrated1994}
  and conductivity \cite{daviesXXXVTransferenceNumbersIonic1924,hamerActivityCoefficientsHydrofluoric1970}.
  This is different for HCl, where excess protons are effectively repelled from chloride ions and are therefore most commonly separated by one or two hydrogen bonds. 
 This difference is also reflected in the theoretically determined excess-proton mobility: relative to water, excess protons diffuse $\sim$3.7 times faster in HCl but only $\sim$1.3 times faster in HF, as shown in the Supplemental Information.
  \\ \\
  A complementary Zundel–Eigen interpretation is applied to HF and HCl solutions and presented in Fig. \ref{fig:HF_HCl_Spectra}(b). 
  In the Zundel-picture the excess proton is shared between two proton acceptors, which can be either water or the conjugate base.
  In HF solutions, $\mathrm{H_3OF}$ configurations dominate (63\%), whereas $\mathrm{H_5O_2^+}$ accounts for only 37\%. 
  In contrast, HCl solutions are overwhelmingly characterized by $\mathrm{H_5O_2^+}$ structures (95\%), with $\mathrm{H_3OCl}$ being only marginally found (5\%).
  These differences reflect the distinct proton solvation preferences in weak and strong acids: fluoride participates directly in proton sharing in a substantial fraction of configurations, whereas chloride plays only a minor role in the corresponding HCl solutions. 
  \\ \\
  In the Eigen picture the excess proton is distributed over four proton acceptors, illustrated by a central hydronium ion (H$_3$O$^+$) hydrogen-bonded to three surrounding molecules, which may be water or the conjugate base. 
  The localization of the positive charge on the hydronium ion in Fig.~\ref{fig:HF_HCl_Spectra}(b) reflects the conventional picture of a hydrogen-bond-stabilized H$_3$O$^+$ ion, whereas in reality these hydrogen bonds can have significant covalent character, especially in the case of fluoride.
  In HCl solutions, the conventional Eigen-cation H$_9$O$_4^+$ is the dominant configuration (72\%), with other Eigen configurations such as H$_7$O$_3$Cl (24\%) and H$_5$O$_2$Cl$_2^-$ (3\%) being less  likely, as shown in Fig. \ref{fig:HF_HCl_Spectra}(b). 
  In HF solutions, the excess proton is found in the Eigen cation H$_9$O$_4^+$ only in 24\% of cases. More commonly, it appears in H$_7$O$_3$F structures (61\%), while 14\% of excess protons are coordinated by two water molecules and two fluoride ions, forming H$_5$O$_2$F$_2^-$ configurations. Further, roughly 1\% of excess protons are coordinated by three fluoride ions (not shown), which leaves only 8\% of fluoride ions uncoordinated by excess protons.
  Because water is present in large excess, adding or removing water molecules from these structures does not change the underlying chemical equilibrium. H$_9$O$_4^+$, H$_7$O$_3$F and H$_5$O$_2$F$_2^-$  can therefore be interpreted as hydrates of H$^+$, HF, and HF$_2^-$, respectively.
  Thermodynamic analysis based on electromotive-force measurements and activity-coefficient estimates predicts H$^+$ and F$^-$ populations of 14\% and 0.1\% in HF solutions at 6~mol/kg \cite{braddyEquilibriaModeratelyConcentrated1994}. Using charge and fluorine balance (see Supplemental Information), this corresponds to 14\%, 72\%, and 14\% for H$^+$, HF, and HF$_2^-$. These values compare well with 24\%, 61\%, and 14\% obtained here at 7.5~mol/kg. Overall, the chemical speciation is consistent, especially given the experimental uncertainties and the modeling approximations discussed in the Supplemental Information, namely, the choice of DFT functional \cite{gillanPerspectiveHowGood2016,brandenburgBenchmarkingDFTSemiempirical2015,calioMinimalExperimentalBias2020,calioResolvingStructuralDebate2021} and the neglect of nuclear quantum effects \cite{napoliDecodingSpectroscopicFeatures2018,atsangoDevelopingMachinelearnedPotentials2023}. \\ \\ 
  We present the simulated absorption coefficients $\tilde{\alpha}(\omega)$ of HF and HCl solutions and water in Fig.~\ref{fig:HF_HCl_Spectra}(d), where $\tilde{\alpha}(\omega)$ denotes the inverse distance over which the intensity of incident light decays to $1/e$ due to absorption \cite{bornPrinciplesOpticsElectromagnetic1999}. The extraction of $\tilde{\alpha}(\omega)$ from AIMD is described in the Supplemental Information and quantitatively tested by comparison between simulated (solid black) and experimental (dashed black) absorption coefficients of pure water \cite{haleOpticalConstantsWater1973b} in Fig.~\ref{fig:HF_HCl_Spectra}(d).
  The agreement is good, with a slight overestimation in the OH-stretch region (3000–3700 cm$^{-1}$), consistent with previous results \cite{brunigTimeDependentFrictionEffects2022}.
  We observe that the absorption coefficients of the HF and HCl solutions are significantly different from pure water.
  Vibrational signatures of excess protons are most clearly identified in rescaled difference spectra,
  \begin{align}
  \frac{\Delta \tilde{\alpha}(\omega)}{X^0_{\mathrm{HA}}}
  &=
  \frac{
  \tilde{\alpha}(\omega)
  -
  X_{\mathrm{H_2O}}\tilde{\alpha}_{\mathrm{H_2O}}(\omega)
  }{
  X^0_{\mathrm{HA}}
  },
  \label{eq:difference_absorb}
  \end{align}
  where $X^0_{\mathrm{HA}}$ is the total mole fraction of acid added to the solution, and $X_{\mathrm{H_2O}}$ is the mole fraction of water molecules excluding the water molecules that form the central hydronium ions in the Eigen picture. $\tilde{\alpha}_{\mathrm{H_2O}}(\omega)$ denotes the absorption coefficient of pure water. As shown in Ref.~\cite{brunigSpectralSignaturesExcessproton2022}, where a similar construction of the difference spectra was employed, spectral signatures defined in this way in HCl solutions vary only weakly with concentration.
  In the simulated and experimental rescaled difference spectra shown in Fig.~\ref{fig:HF_HCl_Spectra}(e), the transfer waiting (TW), transfer path (TP) and normal mode (NM) bands are clearly visible \cite{brunigSpectralSignaturesExcessproton2022}. 

  Here, the TW band reflects the proton-transfer reaction frequency or waiting time, the TP band arises from the proton transfer-path time, and the NM band reflects vibrational oscillations of the proton bound to a base \cite{brunigSpectralSignaturesExcessproton2022} and overlaps with the inverse timescale of the special pair dance\cite{markovitchSpecialPairDance2008}, which is reported to be 15.7~fs in Ref. \cite{calioResolvingStructuralDebate2021}, corresponding to a frequency of 2125~cm$^{-1}$.
  Due to the fundamentally different distributions of Zundel-like structures in HF and HCl solutions presented 
  in Fig. \ref{fig:HF_HCl_Spectra}(b), one would expect significantly different spectra. 
  Rather surprisingly, however, we find that the proton-transfer signatures (TW and TP) are nearly indistinguishable between the two acids, while the normal-mode region exhibits small differences.
  A comparison with experimental difference spectra of HF and HCl solutions in Fig. \ref{fig:HF_HCl_Spectra}(e), which we constructed from absorption spectra reported in the literature \cite{giguereH3OIonsAqueous1976,khoramiEtudeMelangesEau1987,thamerUltrafast2DIR2015} as described in Supplemental Information, shows that HF solutions display pronounced proton-transfer signatures that closely resemble those observed in HCl solutions, in agreement with our theoretical predictions.  
  \\ \\
  In summary, applying the Eigen picture to weak acids yields chemical speciation consistent with thermodynamic predictions, including the debated bifluoride ion. For analyzing spectroscopic signatures of proton transfer, however, the Zundel picture is more useful, 
 since it allows to link the dynamics of the proton-transfer reaction coordinate to IR spectra \cite{brunigSpectralSignaturesExcessproton2022}.
  Given the significant fraction of excess protons in HF coordinated by two water molecules, one might expect similarities between the spectra of HF and HCl solutions. However, the almost indistinguishable spectral proton-transfer signatures are puzzling given that the majority of excess protons are in direct contact with fluoride in HF.
  \subsection*{Free Energy Landscapes of Proton Transfer within H$_5$O$_2^+$ and H$_3$OF}
  \label{sec:free_energy}
  \begin{figure*}
    \centering
    \includegraphics[width=1.\textwidth]{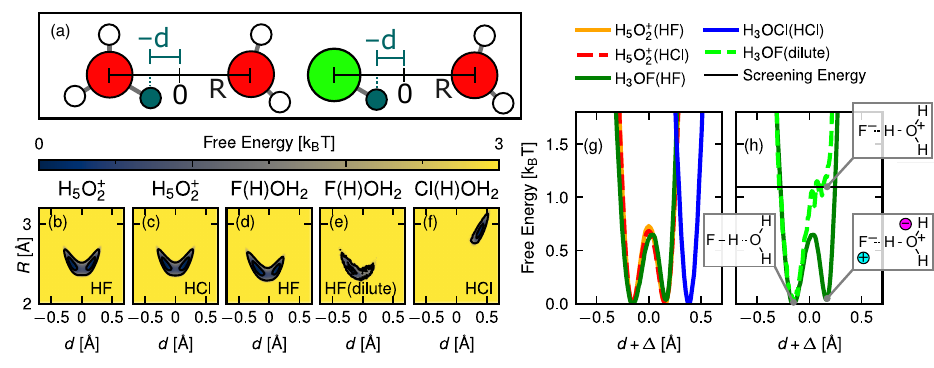}
    \caption{
      Analysis of the excess-proton free-energy landscapes in terms of the proton-transfer coordinate $d$ and the heavy-atom-distance $R$, as schematically 
      illustrated in (a): the
      coordinate $d$ is the projection of the excess-proton position onto the heavy-atom connecting vector
      axis (O--O, F--O, or Cl--O) of length $R$.
      The excess proton is colored in dark turquoise and in all cases, negative $d$ correspond to the proton being closer to the left heavy atom.
      (b--f) Two-dimensional free-energy surfaces of the excess proton for the Zundel
      configuration: H$_5$O$_2^+$ in HF, H$_5$O$_2^+$ in HCl, H$_3$OF in HF, H$_3$OF in dilute HF, and H$_3$OCl in HCl,
      respectively.
      (g) One-dimensional free-energy profiles along the proton-transfer coordinate $d$, for the Zundel configuration in concentrated HF and HCl solutions.
  (h) Same as in (g), but only for H$_3$OF in concentrated and in dilute HF to highlight the effect of screening stabilization of the ion pair F$^- \cdots$H$_3$O$^+$ in HF.
  The free-energy shift of the ion pair F$^- \cdots$H$_3$O$^+$ of 1.1~$k_\mathrm{B}T$ between the two concentrations is indicated by the horizontal line. The mechanism underlying the energetic similarity between H$_5$O$_2^+$ and H$_3$OF in HF is illustrated schematically by the chemical structures and surrounding ions shown in the insets of (h).
  To facilitate comparison with H$_5$O$_2^+$, the $d$ coordinate of H$_3$OF is shifted by $\Delta = 0.04\,\text{\AA}$ in (g) and (h). 
      }
    \label{fig:free_energy}
  \end{figure*}
  We analyze the free energy of the excess proton as a function of the coordinates $R$ and $d$, illustrated in Fig.~\ref{fig:free_energy}(a). Here, $R$ denotes the distance between the heavy atoms and the proton-transfer coordinate $d$ is defined as the projection of the excess proton position onto the heavy-atom connecting axis (O--O, F--O, or Cl--O). 
  \\ \\ 
  As can be seen by comparison of Figs. \ref{fig:free_energy}(b-d) the two-dimensional free-energy surfaces of H$_5$O$_2^+$ in HF,  H$_5$O$_2^+$ in HCl and H$_3$OF in HF are very similar, with the most notable difference being a slight shift of the excess proton towards fluoride (i.e. to the left) for large heavy-atom distances $R$ in H$_3$OF.
  All three free-energy surfaces exhibit the characteristic double-well shape for large heavy-atom distances $R$ and a single-well shape for small $R$, which is typical for Zundel configurations and is qualitatively consistent with previous works 
  \cite{marxNatureHydratedExcess1999,sillanpaaStructuralSpectralProperties2002,simonInitioMolecularDynamics2005,marxProtonTransfer2002006,brunigSpectralSignaturesExcessproton2022}.
  The striking similarity between the free-energy surfaces of H$_5$O$_2^+$ (Figs.~\ref{fig:free_energy}(b,c)) and H$_3$OF in HF (Fig.~\ref{fig:free_energy}(d)) explains the nearly identical proton-transfer signatures observed in the IR spectra of HF and HCl solutions.
  \\ \\
In the H$_3$OCl configuration the proton is strongly localized towards the water molecule, as shown in Fig. \ref{fig:free_energy}(f), which is a consequence of the negligible proton affinity of chloride. In order to understand the underlying mechanism for the energetic similarity between H$_5$O$_2^+$ and H$_3$OF in HF, we additionally simulated a single HF molecule in water using AIMD and computed the free-energy surface of H$_3$OF in dilute HF, shown in Fig.~\ref{fig:free_energy}(e).
  As can be seen by comparison of Figs.~\ref{fig:free_energy}(d,e), the proton in H$_3$OF in dilute HF is localized slightly more towards the fluoride ion, while in concentrated HF it is almost equally shared between the water molecule and the fluoride ion, with a free-energy landscape comparable to that of H$_5$O$_2^+$.
  \\ \\
  To enable a more quantitative comparison between the different Zundel configurations, we present one-dimensional free-energy profiles along the proton-transfer reaction coordinate $d$, obtained by integrating the two-dimensional probability distribution over $R$, which we show in Figs.~\ref{fig:free_energy}(g,h).
  Approximately, the reaction coordinate $d$ describes the proton transfer reaction according to
  \begin{align}
  \mathrm{AH} \cdots \mathrm{OH_2}
  \rightleftharpoons
  \mathrm{A}^- \cdots \mathrm{H_3O}^+ , 
  \label{eq:three_step_scheme}
  \end{align}
  with $A \in \lbrace \mathrm{H_2O^+}, \mathrm{F}, \mathrm{Cl} \rbrace$, where the dots indicate hydrogen bonds. The state $\mathrm{AH} \cdots \mathrm{OH_2}$ ($d<0$) corresponds to the proton bound to $A^-$, and $\mathrm{A}^- \cdots \mathrm{H_3O}^+$ ($d> 0$) is the configuration where the excess proton is localized on the water molecule.
  As can be seen in Fig. \ref{fig:free_energy}(g), the
   $\Delta$-shifted one-dimensional free-energy profiles of H$_5$O$_2^+$ in HF (solid orange), H$_5$O$_2^+$ in HCl (dashed red) and H$_3$OF (solid dark green) in HF are almost indistinguishable, with barriers around 0.7~$k_B T$. 
  In contrast, the free-energy profile of H$_3$OCl in HCl (solid blue) is highly asymmetric, exhibiting a single well localized toward the water molecule, consistent with the negligible proton affinity of chloride.
  The free-energy profiles of H$_3$OF in concentrated HF (solid dark green) and in dilute HF (dashed lime green) are shown in Fig.~\ref{fig:free_energy}(h) to highlight the effect of screening on the stabilization of the ion pair F$^- \cdots$H$_3$O$^+$. 
  In dilute HF, H$_3$OF exhibits a single well localized toward the fluoride ion. In concentrated HF, however, the ion pair is stabilized by the surrounding ionic environment, resulting in a free-energy lowering of approximately $1.1\,k_B T$, as indicated by the horizontal line in Fig.~\ref{fig:free_energy}(h).
  \\ \\ 
  The free energy stabilization of two ions F$^-$ and H$_3$O$^+$ at separation $R$ due to ionic screening is predicted by Debye-Hückel theory \cite{Debye1923} to be
  \begin{align}
  \Delta F = -\frac{e^2}{4\pi \varepsilon_0 \varepsilon} \frac{\kappa}{1+\kappa R} = -1.5~k_\mathrm{B}T,
  \label{eq:screening_energy}
  \end{align}
  in decent agreement with the observed free-energy difference of $-1. 
  1~k_\mathrm{B}T$ between the 6.5~mol/L and the dilute solution in Fig.~\ref{fig:free_energy}(h).
  Here, $e$ is the elementary charge, $\varepsilon_0$ is the 
  vacuum permittivity,
  $\kappa^{-1}=~2.4~\text{\AA}$ is the Debye length at concentration of free ions of $c=24\%\times 6.5~\mathrm{mol\,L^{-1}}$, $R=2.4~\text{\AA}$ is the mean distance between the hydronium and the fluoride ion (compare Figs. \ref{fig:free_energy}(d,e)) and $\varepsilon=78$ is the dielectric constant of water \cite{haynesCRCHandbookChemistry2015a}. 
   Since the equilibrium in Eq.~\eqref{eq:three_step_scheme} remains unchanged for H$_5$O$_2^+$ due to symmetry but shifts as a function of ionic strength for H$_3$OF, one expects a stronger concentration dependence of proton-transfer spectra in HF than in HCl, where it was found to be minimal \cite{brunigSpectralSignaturesExcessproton2022}.
   \\ \\
  \section*{Conclusion}
  Our analysis reconciles thermodynamic and spectroscopic properties of aqueous HF and provides a consistent picture of excess proton solvation in the presence of conjugate bases with significant proton affinity, such as F$^-$. An extended Eigen framework yields chemical speciation consistent with thermodynamic predictions, including the elusive bifluoride ion.
  Despite distinct excess-proton structures found in HF and HCl, the proton-transfer bands are nearly identical. 
  This apparent contradiction is resolved by screening due to free ions, which renders the free-energy profiles of the proton transfer coordinate within the relevant Zundel configurations nearly indistinguishable in HF and HCl, leading to convergent proton-transfer dynamics and spectroscopic signatures.
  These free-energy profiles and spectral signatures are connected through the generalized Langevin equation \cite{brunigTimeDependentFrictionEffects2022,colinetNonMarkovianLinearVibrational2025}. 
  Our complementary Eigen-Zundel interpretation is crucial: Eigen-like structures determine equilibrium populations, while Zundel-like configurations encode the transition states that govern vibrational response. We expect that our findings are applicable to weak acids in general. However, we suggest that the small size of F$^-$ enables the formation of H$_5$O$_2$F$_2^-$, which would be sterically hindered for larger anions and therefore explains the anomalous thermodynamic behavior of HF compared to other weak acids. 
  \clearpage

  \section*{Acknowledgements}
  We acknowledge financial support from the Deutsche Forschungsgemeinschaft (DFG) within the CRC 1349. Computational resources were provided by the HPC systems CURTA \cite{bennettCurtaGeneralpurposeHighPerformance2020}, TRON, and Sheldon at Freie Universität Berlin, as well as by the North-German Supercomputing Alliance (HLRN) under project bep00068.

\section*{Conflict of Interest}

The authors declare no conflict of interest.


\bibliography{mybib}
\end{document}


\title[]{Supplemental Information}

\author{Louis Lehmann}
\affiliation{Department of Physics, Freie Universität Berlin, Arnimallee 14, 14195 Berlin, Germany}

\author{Florian N. Brünig}
\affiliation{Department of Physics and Materials Science, University of Luxembourg, L-1511 Luxembourg City, Luxembourg}

\author{Jonathan Scherlitzki}
\affiliation{Department of Physical and Theoretical Chemistry, Freie Universität Berlin, Arnimallee 22, 14195 Berlin, Germany}

\author{Morten Lehmann}
\affiliation{Institute of Chemistry, Technische Universität Berlin, Straße des 17. Juni 115, 10623 Berlin, Germany}

\author{Martin Kaupp}
\affiliation{Institute of Chemistry, Technische Universität Berlin, Straße des 17. Juni 115, 10623 Berlin, Germany}

\author{Beate Paulus}
\affiliation{Department of Physical and Theoretical Chemistry, Freie Universität Berlin, Arnimallee 22, 14195 Berlin, Germany}

\author{Roland R. Netz$^*$}
\email{corresponding author: rnetz@physik.fu-berlin.de}
\affiliation{Department of Physics, Freie Universität Berlin, Arnimallee 14, 14195 Berlin, Germany}

\maketitle
\tableofcontents
\newpage
\section{Computational Details}
\label{sec:computational_details}
All production simulations were performed in the NVT ensemble at 300 K using CP2K \cite{kuhneCP2KElectronicStructure2020a}. The BLYP exchange-correlation functional \cite{beckeDensityfunctionalExchangeenergyApproximation1988,leeDevelopmentColleSalvettiCorrelationenergy1988} with D3 dispersion corrections \cite{grimmeConsistentAccurateInitio2010} was employed. 
Core electrons were treated with Goedecker-Teter-Hutter (GTH) pseudopotentials \cite{goedeckerSeparableDualspaceGaussian1996a}. DZVP-MOLOPT-SR-GTH basis sets were used for O and H, and aug-DZVP-GTH for F and Cl \cite{vandevondeleGaussianBasisSets2007a}. A plane-wave cutoff of 400 Ry was applied. The timestep was 0.5 fs, and temperature control was achieved using a CSVR thermostat \cite{bussiCanonicalSamplingVelocity2007} with a 100 fs time constant.
Simulations contained 256 water molecules for pure water and 224 water molecules for the 7.5~mol/kg acid solutions, with 30 excess protons and 30 anions in the latter. The pure water system was prepared as described in Ref. \cite{brunigTimeDependentFrictionEffects2022}. Initial HCl structures were generated by classical MD using GROMACS \cite{abrahamGROMACSHighPerformance2015a}: 200~ps in the NPT ensemble at ambient pressure followed by 800~ps in the NVT ensemble at 300 K, employing the SPC/E water model \cite{berendsenMissingTermEffective1987} and Cl$^-$ and H$_3$O$^+$ force-field parameters from Ref. \cite{bonthuisOptimizationClassicalNonpolarizable2016}. For HF, anions were replaced and configurations rescaled according to experimental density ratios \cite{cleggDensitiesApparentMolar2011,AcidBaseNormality}. The first 14 ps (HF) and 9 ps (HCl) were discarded for equilibration.
The amount of Zundel- and Eigen structures are found to be in chemical equilibrium after this time.
Box lengths were 19.73 \AA~ (pure water), 20.23 \AA~ (7.5~mol/kg HCl), and 19.63 \AA~ (7.5~mol/kg HF). Trajectory lengths were 201~ps (water), 91~ps (HCl), and 128~ps (HF). Water and HCl trajectories were previously analyzed \cite{brunigTimeDependentFrictionEffects2022,brunigSpectralSignaturesExcessproton2022} and are reused here for comparison with HF.
To simulate a dilute HF solution, a single HF molecule in water was simulated in a 9.855~\AA~cubic box containing 31 water molecules. Therefore, a stable configuration of 32 water molecules at experimental density (1.0 g/cm$^{3}$) was generated via a classical 20~ps NVT simulation as described above (but with a reduced Lennard-Jones cutoff of 4.5~\AA), and then equilibrated for 10~ps with AIMD after replacing one water molecule by HF. The remaining 39~ps were used for analysis. 
\\ \\ 
The IR spectra were calculated from the Fourier transformation of the polarization autocorrelation function using Wannier center localization 
which is briefly described in Ref. \cite{brunigSpectralSignaturesExcessproton2022} and detailed in Sections \ref{sec:absorption_coefficient} and \ref{sec:dielectric_spectrum_from_aimd}. 
 \\
Bound protons are defined as the two protons closest to each oxygen atom in the system, the remaining protons are identified as excess protons. Zundel configurations ($\mathrm{H_5O_2^+}$, $\mathrm{H_3OF}$, and $\mathrm{H_3OCl}$) are identified based on the two closest heavy atoms (O, F, or Cl) to an excess proton. 
A similar, but slightly more complex definition of excess protons was used in Ref. \cite{brunigSpectralSignaturesExcessproton2022}, which was needed to obtain undisrupted trajectories of excess protons for the spectral analysis of the barrier-crossing dynamics. 
Eigen configurations are defined by a central H$_3$O$^+$. The H$_3$O$^+$ oxygen is taken as the oxygen atom closest to the excess proton.
Proton sharing anions are defined as the hydrogen bonded neighbors to the H$_3$O$^+$ oxygen. 
Hydrogen bonds are defined by the geometric definition of Ref. \cite{ozkanlarUnifiedDescriptionHydrogen2014}. 
To enforce mass and charge conservation, each anion (F$^-$, Cl$^-$) is assigned to its nearest H$_3$O$^+$, avoiding double counting. Consequently, H$_2$O molecules in the Eigen-picture (Fig.~1(b) of the main text) should be understood as generic hydrogen-bond acceptors, which are mostly water but can also be an anion bound to a different H$_3$O$^+$.
Counting hydrogen-bond partners of each hydronium ion yields the distributions: HF (H$_9$O$_4^+$ 17\%, H$_7$O$_3$F 44\%, H$_5$O$_2$F$_2^-$ 28\%) and HCl (H$_9$O$_4^+$ 64\%, H$_7$O$_3$Cl 26\%, H$_5$O$_2$Cl$_2^-$ 3\%), which differ from those reported in the main text, since only unique anion assignments to H$_3$O$^+$ ions are used there.
 Since Braddy et al.~\cite{braddyEquilibriaModeratelyConcentrated1994} report only the molalities of H$^+$ and F$^-$, the molalities of HF and HF$_2^-$ in the experimental data are obtained from charge conservation, $m_{\mathrm{H}^+}=m_{\mathrm{HF}_2^-}+m_{\mathrm{F}^-}$, and fluorine conservation, $m^0_{\mathrm{HF}}=m_{\mathrm{HF}}+m_{\mathrm{F}^-}+2\times m_{\mathrm{HF}_2^-}$, where $m_\alpha$ denotes the molality of species $\alpha$ and $m^0_{\mathrm{HF}}$ the initial HF molality. This treatment effectively reduces the analysis to the building blocks H$^+$, HF, F$^-$, and HF$_2^-$
and does not consider H$_2$F$_3^-$ formation, which was considered in the analysis of Ref. \cite{braddyEquilibriaModeratelyConcentrated1994}.
This is consistent with our simulation analysis, where we do not investigate aggregation of different Eigen-like structures.
The shortest path between the excess proton and the conjugate base is determined by the breadth-first-search algorithm, using the implementation in the Python package NetworkX \cite{hagbergExploringNetworkStructure2008}. We neglect nuclear quantum effects and therefore compute the free-energy landscapes from classical probability distributions. Although nuclear quantum effects can modify these distributions \cite{schmittComputerSimulationProton1999,marxProtonTransfer2002006,atsangoDevelopingMachinelearnedPotentials2023}, they are expected to affect all systems similarly and thus are not expected to alter the presented comparison. Their influence on IR spectra is also small, since classical and quantum harmonic oscillators share the same position response function \cite{kuboFluctuationdissipationTheorem1966,caldeiraQuantumTunnellingDissipative1983}, which explains why no quantum correction factor \cite{ramirezQuantumCorrectionsClassical2004} needs to be applied to the IR spectra in this work, as described in the Supplemental Information of Ref. \cite{lehmannMultipolarElectricMagnetic2026}.
\section{Discussion of the Employed Level of Theory and Quantum Chemistry Calculations}
\label{sec:level_of_theory}
\subsection*{Assessment of BLYP-D3 Based on Literature}
The here employed BLYP-D3 level of theory provides a practical compromise between accuracy and computational cost. In a broad benchmark of density functionals for water properties \cite{gillanPerspectiveHowGood2016}, it achieved an overall score of 66/100, outperforming methods such as BLYP (42/100),  PBE (56/100), PBE-TS (56/100) \cite{perdewGeneralizedGradientApproximation1996,tkatchenkoAccurateMolecularVan2009}
and revPBE-DRSLL (49/100) \cite{dionVanWaalsDensity2004,zhangCommentGeneralizedGradient1998}, although optPBE-DRSLL and optB88-DRSLL performed better (74/100) \cite{dionVanWaalsDensity2004,klimesChemicalAccuracyVan2009}. 
BLYP-D3 has also been shown to accurately reproduce spectral properties of pure water \cite{brunigSpectralSignaturesExcessproton2022} and HCl solutions at various concentrations \cite{brunigTimeDependentFrictionEffects2022} and is demonstrated to give excellent agreement with the lattice energy of ice polymorphs \cite{brandenburgBenchmarkingDFTSemiempirical2015}.
Nevertheless, BLYP-D3 has known limitations. It substantially underestimates the experimental water self-diffusion coefficient $D= 0.23\,\text{\AA}^2/\text{ps}$ \cite{harrisPressureTemperatureDependence1980}, yielding $D= 0.07\pm0.01\,\text{\AA}^2/\text{ps}$ in the dataset analyzed here, consistent with previous studies \cite{calioMinimalExperimentalBias2020}. It also underestimates the proton-transfer barrier in isolated Zundel cations relative to CCSD(T) \cite{calioResolvingStructuralDebate2021}.
Both drawbacks can be corrected using an additional atomic pair potential 
(EDS-BLYP-D3)
\cite{calioMinimalExperimentalBias2020,calioResolvingStructuralDebate2021}, which is not parameterized for oxygen-fluoride and hydrogen-fluoride interactions and was therefore not used here. In summary, the literature reports different assessments for the ability of BLYP-D3 to accurately describe aqueous systems.
In the following section we report additional quantum-chemical calculations of reaction energies in the Zundel- and Eigen-picture to assess the applicability of BLYP-D3 for the systems studied here.
\subsection*{Quantum Chemistry Calculations of Reaction Energies}
\begin{figure}
  \centering
  \includegraphics[width=0.8\textwidth]{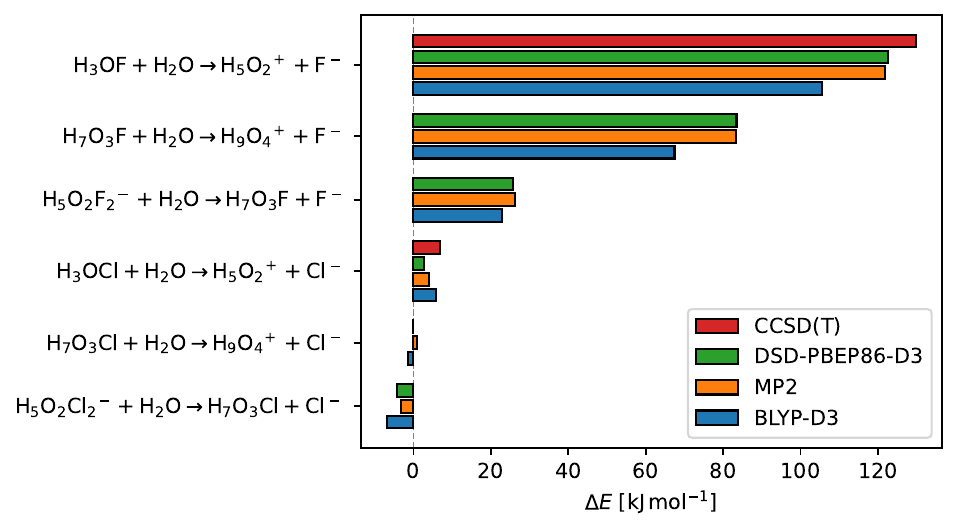}
  \caption{Reaction energies $\Delta E$ of the proton-transfer reactions in the Zundel- and Eigen-picture for different levels of theory, based on quantum chemistry calculations of the energy-optimized structures sketched in Fig.~1(b) of the main text.
  The reaction equations are shown on the y-axis. The energies are computed using CCSD(T) (red), MP2 (orange), DSD-PBEP86-D3 (green) and BLYP-D3 (blue). 
  Note, that the reaction energies are not directly comparable to experimental values since the solvent is only taken into account in a rather approximative manner and only the minimum-energy geometry of each structure is considered. 
  }
  \label{fig:level_of_theory_comparison}
\end{figure}
Here, we present reaction energies obtained from quantum-chemical calculations involving the Eigen- and Zundel-like structures depicted in Fig.~1(b) of the main text.  We note that the reaction energies obtained from these calculations are not directly comparable to experimental values for two reasons: Firstly, the solvent environment is treated only approximately using the conductor-like polarizable continuum model (C-PCM) \cite{Barone1998} and, secondly, only the minimum-energy geometry of each structure is considered. Nevertheless, these calculations provide a quantitative assessment of the accuracy of the BLYP-D3 level of theory and allow us to draw qualitative conclusions about the relative stability of the different Zundel- and Eigen-like structures observed in the bulk AIMD simulations. We used ORCA 6.1.0 \cite{Neese2020, Neese2022} to perform geometry optimizations of the water molecule \ce{H2O} and of the Zundel cation \ce{H5O2+}, the Eigen cation \ce{H9O4+}, as well as \ce{H3OX}, \ce{H7O3X}, and \ce{H5O2X2-}, with $\ce{X} \in {\ce{F},\ce{Cl}}$. For \ce{H} and \ce{O}, we employed the cc-pVTZ basis set \cite{Dunning1989}, whereas for \ce{F} and \ce{Cl} we used aug-cc-pVTZ \cite{Dunning1989, Kendall1992, Woon1993}. We note that the comparatively small basis set used in the bulk AIMD simulations constitutes an additional source of error that is not considered in the following analysis. We repeated these calculations at four different levels of theory using tight convergence thresholds for the self-consistent field energies and gradients. The considered levels of theory are BLYP-D3, consisting of BLYP DFT with Grimme's D3 dispersion correction \cite{Grimme2010, Grimme2011} and Becke--Johnson damping \cite{Becke2005, Johnson2005}, MP2, based on second-order M{\o}ller--Plesset perturbation theory \cite{Moller1934} using the resolution-of-the-identity (RI) approximation, DSD-PBEP86-D3, employing the double-hybrid DSD-PBEP86 functional with D3 dispersion and Becke--Johnson damping \cite{Kozuch2011, Kozuch2013, Goerigk2017} and RI for the MP2 correlation contribution, and CCSD(T), including coupled-cluster singles, doubles, and perturbative triples \cite{Purvis1982} with numerical gradients. CCSD(T) serves as the reference method. However, because of its high computational cost, for the larger systems MP2 and DSD-PBEP86-D3 are used as more affordable alternatives.

We define $\Delta E=E_\mathrm{B}-E_\mathrm{A}$ for the reaction $\mathrm{A}\rightarrow\mathrm{B}$, where $E_\mathrm{A}$ and $E_\mathrm{B}$ are the internal energies of the reactant and product states A and B, respectively. The proton-transfer energy is computed by replacing the conjugate base in each proton-bearing species with a water molecule. The corresponding reaction equations are given on the y-axis of Fig.~\ref{fig:level_of_theory_comparison}. We consider the Zundel-like and Eigen-like structures depicted in Fig.~1(b) of the main text and compare the resulting proton-transfer energies obtained at the different levels of theory in Fig.~\ref{fig:level_of_theory_comparison}. Overall, BLYP-D3 reproduces the energy differences obtained at the other levels of theory reasonably well, as shown by the agreement between the BLYP-D3 results (blue bars) and those obtained with CCSD(T) (red), MP2 (orange), and DSD-PBEP86-D3 (green). However, BLYP-D3 underestimates the binding energy of an excess proton to fluoride, which suggests that the predicted equilibrium between the Zundel- and Eigen-like structures shown in Fig.~1(b) of the main text likely underestimates the fraction of excess protons coordinated to fluoride. Coordinates of the optimized structures are provided in Tables \ref{tab:BLYP-water}-\ref{tab:CCSD(T)-zundel-cl}.
In the following, the individual proton-transfer reactions are discussed briefly. The first reaction considered is proton transfer in the Zundel picture
\begin{equation}
\mathrm{H_3OX} + \mathrm{H_2O} \rightarrow \mathrm{H_5O_2^+} + \mathrm{X}^- \, .
\label{eq:proton_transfer_zundel}
\end{equation}
For $\mathrm{X}^-=\mathrm{F}^-$, the proton-transfer energy is strongly positive at all levels of theory, whereas for $\mathrm{X}^-=\mathrm{Cl}^-$ it is substantially smaller, which is consistent with the thermodynamic expectation that HF is a much weaker acid than HCl.
The same observation can be made for the proton-transfer reaction in the Eigen picture, which differs from the Zundel picture by the presence of two additional water molecules in the proton-bearing species
\begin{equation}
\mathrm{H_7O_3X} + \mathrm{H_2O} \rightarrow \mathrm{H_9O_4^+} + \mathrm{X}^- \, .
\label{eq:proton_transfer_eigen}
\end{equation}
Note, that the reaction in the Eigen-picture in Eq. \eqref{eq:proton_transfer_eigen} and in the Zundel-picture in Eq. \eqref{eq:proton_transfer_zundel} both describe the same reaction, but only differ in the number of water molecules explicitly included in the quantum-chemical calculations. Interestingly, the reaction energy is smaller for both anions in the Eigen picture than in the Zundel picture.
Since the dielectric contribution of the aqueous solvent is already accounted for through C-PCM, this additional stabilization must originate from explicit chemical interactions with the surrounding water molecules. This, in turn, suggests that the Eigen picture provides a more appropriate description of the equilibrium structure of the hydrated excess proton, in agreement with Ref.~\cite{calioResolvingStructuralDebate2021} and our interpretation of the bulk AIMD simulations.
Finally, we consider the ligand exchange energy for the Eigen-like difluoride \ce{H5O2F2-} and dichloride \ce{H5O2Cl2-} anions, in which one anion is replaced by a water molecule to form the corresponding Eigen-like structure containing a single anion
\begin{equation}
\mathrm{H_5O_2X_2^-} + \mathrm{H_2O} \rightarrow \mathrm{H_7O_3X} + \mathrm{X}^- \, .
\label{eq:proton_transfer_bifluoride}
\end{equation}
We find that an excess proton is strongly stabilized by an additional fluoride ion, whereas the stabilization provided by an additional chloride ion is substantially weaker. This result indicates that a single excess proton can be stabilized by multiple fluoride ions, consistent with our bulk AIMD findings presented in the main text.
\\
\section{Excess Proton Diffusion in HF and HCl Solutions}
\begin{figure}
  \centering
  \includegraphics[width=1.\textwidth]{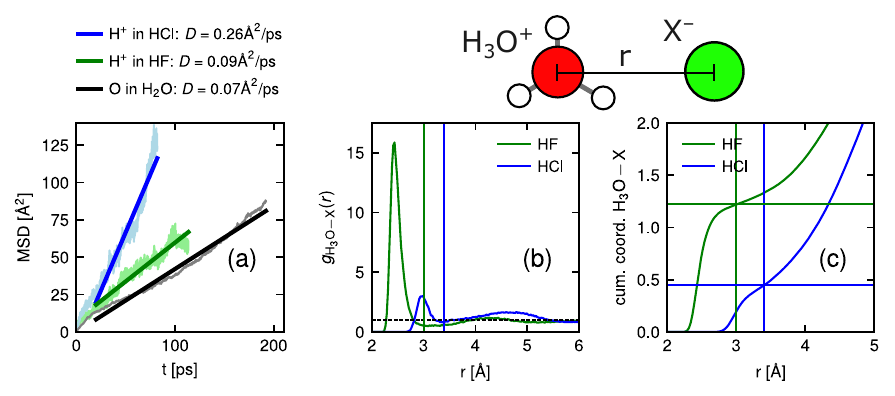}
  \caption{(a): Mean-square displacement (MSD) of excess protons in the 7.5~mol/kg HF (green) and HCl (blue) solutions as a function of time, compared with the MSD of water molecules in pure water (black). The dark lines show linear fits to the MSD data for $t>20,\text{ps}$, from which the corresponding diffusion coefficients are determined. The lighter lines show the calculated MSD trajectories. (b): Radial distribution function between hydronium oxygen atoms and the conjugate base anions F$^-$ (green) and Cl$^-$ (blue) in the HF and HCl solutions, respectively. (c): Cumulative coordination number of the conjugate base anions around hydronium oxygen atoms in the HF and HCl solutions, obtained by integrating the radial distribution functions shown in panel (b). The vertical lines in panels (b) and (c) indicate the first coordination shell and the horizontal lines in panel (c) indicate the corresponding coordination numbers.
  }
  \label{fig:diffusion_coefficients}
\end{figure}
Here, we describe the calculation of the diffusion constants of excess protons in the concentrated HF and HCl solutions from the AIMD simulations and interpet the results.
 The diffusion coefficient $D$ is obtained from the mean-square displacement (MSD) of the excess protons according to
\begin{equation}
  D = \lim_{t \to \infty} \frac{1}{6 t} \langle |\vec r(t) - \vec r(0)|^2 \rangle \, ,
\end{equation}
\cite{einsteinUberMolekularkinetischenTheorie1905}, where $\vec r(t)$ is the position and $\langle \cdot \rangle$ denotes an ensemble average. 
We define excess protons as described in Section \ref{sec:computational_details}. To determine the position of an excess proton, we assign it to the closest heavy atom (O, F, or Cl). We then track the motion of the excess proton by following the heavy atom to which it is associated, as long as no proton transfer occurs. 
To identify proton transfer, we monitor all protons covalently bound to this heavy atom. When one of these protons is transferred to another heavy atom, the receiving heavy atom becomes the new reference site, and the trajectory of the excess proton is continued from there. In rare cases, simultaneous proton-transfer events occur, such that the receiving heavy atom immediately transfers a proton to a third heavy atom and therefore no longer carries the excess charge. In these cases, we follow the subsequent receiving heavy atom and continue the excess-proton trajectory from that site. This implementation is similar to the one in Ref. \cite{brunigSpectralSignaturesExcessproton2022}. Our results for HCl and H$_2$O are consistent with those in Ref. \cite{brunigSpectralSignaturesExcessproton2022}, but HF was not considered in that work. 
As in Ref. \cite{brunigSpectralSignaturesExcessproton2022}, we fit the mean-square displacement (MSD) to a linear function for $t>20\,\mathrm{ps}$ to determine the diffusion coefficient. 
These fits alongside the time dependend MSD are shown in Fig.~\ref{fig:diffusion_coefficients}a. 
We find that the diffusion coefficient of excess protons in the HF solution is $D^\mathrm{H^+}_\mathrm{HF} = 0.09\,\text{\AA}^2/\text{ps}$, which is way smaller than in the HCl solution $D^\mathrm{H^+}_\mathrm{HCl} = 0.26 \,\text{\AA}^2/\text{ps}$. For reference, we obtain a water diffusion coefficient of $D^{\mathrm{H_2O}}_\mathrm{H_2O}=0.07\,\text{\AA}^2/\text{ps}$, substantially smaller than the experimental value as already mentioned in Section \ref{sec:level_of_theory}. 
Calio \textit{et al.} used temperature-dependent radial distribution functions to elucidate the molecular origins of proton-transfer barriers in HCl solutions \cite{calioMolecularOriginsBarriers2020}. Direct comparison is not possible because our simulations are available only at a single temperature, but the solvation structure can still be characterized through the radial distribution function between hydronium oxygen atoms and the conjugate-base anions, shown in Fig.~\ref{fig:diffusion_coefficients}(b). Here, ''hydronium oxygen'' denotes the oxygen atom closest to the excess proton, even when a conjugate-base anion lies closer to the proton, as frequently occurs in HF solutions. Figure~\ref{fig:diffusion_coefficients}(b) reveals pronounced differences in the hydronium solvation structure between HF and HCl solutions.
 The first peak of the H$_3$O$^+$-F$^-$ radial distribution function is located at $2.4\,\text{\AA}$, which is significantly smaller than the corresponding H$_3$O$^+$-Cl$^-$ distance at $3\,\text{\AA}$. The radial distribution of the hydronium oxygen and Cl$^-$ agrees well with the one reported by Calio et al. \cite{calioMolecularOriginsBarriers2020}.
 Further, the peak height of the H$_3$O$^+$-F$^-$ radial distribution function is substantially larger than that of the H$_3$O$^+$-Cl$^-$ radial distribution function. Together, these observations demonstrate that excess protons are strongly coordinated to fluoride ions, whereas chloride ions are only weakly coordinated to excess protons, in agreement with thermodynamic expectation and the quantum-chemical calculations of the reaction energies in Section \ref{sec:level_of_theory}.
 By integrating the radial distribution functions, we obtain the cumulative coordination number of the conjugate base anions around hydronium oxygen atoms, which is shown in Fig.~\ref{fig:diffusion_coefficients}(c). We find that within the first solvation shell of hydronium oxygen atoms, there are on average 1.22 fluoride ions and only 0.45 chloride ions. 
Together, these results demonstrate that excess protons are strongly coordinated to fluoride ions, which explains why the mobility of excess protons in HF solutions is significantly lower and demonstrates that our AIMD simulations capture the expected thermodynamic behavior of weak and strong acid solutions.
\color{black}
\section{Relation Between the Absorption Coefficient and the Dielectric Spectrum}
\label{sec:absorption_coefficient}
We consider the one-dimensional inhomogeneous wave equation in vacuum 
\begin{equation}
  \left( \frac{\partial^2}{\partial z^2} + \frac{\omega^2}{c^2} \right) E(z,t) = \mu_0 \dot{J}(z,t) ,
  \label{eq:wave_equation}
  \end{equation}
with a current density $J(z,t)$ induced by a sinusoidal electric field given by  $E(z,t)=\mathcal{E}(z) e^{-i \omega t}+c.c.$. 
Here $c$ is the speed of light in vacuum, $\mu_0$ the vacuum permeability 
\cite{bornPrinciplesOpticsElectromagnetic1999}, $c.c.$ is the complex conjugate and $\mathcal{E}(z)$ the z-dependent amplitude of the electric field.
 If the electric field is sufficiently weak, the current density is related to the electric field by the constitutive relation 
\begin{equation}
J(z,t) = \varepsilon_0 \tilde{\sigma_\mathrm{P}}(\omega)  \mathcal{E}(z) e^{-i \omega t}+c.c. \, ,
\label{eq:constitutive_relation_J}
\end{equation}
where $\tilde{\sigma_\mathrm{P}}(\omega)$ is a complex frequency-dependent conductivity and $\varepsilon_0$ is the vacuum permittivity. 
We associate all charges in our system to the polarization density $P(z,t)$ which obeys
\begin{equation}
\dot{P}(z,t)=J(z,t) \, .
\label{eq:current_density_polarization}
\end{equation}
We note that the conductivity $\tilde{\sigma_\mathrm{P}}(\omega)$ is the polarization conductivity and is thus different from the usual conductivity, which only accounts for transport of free charges in response to an electric field \cite{jacksonClassicalElectrodynamicsInternational2021}. Introducing the complex dielectric function as 
\begin{equation}
  \omega \tilde{\varepsilon}(\omega) = \omega + i \tilde{\sigma}_\mathrm{P}(\omega) \, ,
  \label{eq:dielectric_function}
  \end{equation}
  we obtain from Eqs \eqref{eq:constitutive_relation_J} and \eqref{eq:current_density_polarization}
  \begin{equation}
    J(z,t)=\dot P (z,t) = -i \omega \varepsilon_0 \left( \tilde{\varepsilon}(\omega) - 1 \right) \mathcal{E}(z) e^{-i \omega t}+c.c. \, .
  \end{equation}
With this, we eliminate the current density in Eq.~\eqref{eq:wave_equation} and obtain the homogeneous wave equation in matter
\begin{equation}
  \left( \frac{\partial^2}{\partial z^2} + \tilde{k}(\omega)^2 \right) E(z,t) = 0 \, ,
  \end{equation}
  where we introduced the complex wave vector $\tilde{k}(\omega) = \frac{\omega}{c}\sqrt{\tilde{\varepsilon}(\omega)}$ and used the relation $c^2 = 1/(\varepsilon_0 \mu_0)$ \cite{bornPrinciplesOpticsElectromagnetic1999}.
The solution for the electric field is given by
\begin{equation}
  E(z,t) = \mathcal{E}_0 e^{-i \left( \omega t \pm \tilde{k}(\omega) z \right)} + c.c. \, ,
\end{equation}
where the sign in the exponent determines the direction of propagation and $\mathcal{E}_0$ is the z-independent amplitude. The absorption coefficient $\tilde \alpha(\omega)$ is defined as the inverse length at which the intensity of the electric field decreases by a factor $1/e$ and is thus given by 
\begin{align}
  \tilde \alpha(\omega) &= 2 \tilde{k}(\omega)'' \, ,
  \label{eq:absorption_coefficient}
\end{align}
where $\tilde{k}(\omega)''$ is the imaginary part of the complex wave vector. The factor $2$ appears because the intensity is proportional to the square of the electric field. 
\section{Obtaining the Dielectric Spectrum from Ab-Initio Molecular Dynamics Simulations} 
\label{sec:dielectric_spectrum_from_aimd}
We consider an overall charge-neutral system of $N$ atoms with charges $q_i$, and  electric dipole moments $\mu_i(t)$, occupying a volume $V$ under periodic boundary conditions.  
The current density is given by \cite{jacksonClassicalElectrodynamicsInternational2021}
\begin{equation}
  J(t) = \frac{1}{V} \sum_{i=1}^{N} \left( q_i v_i(t) + \dot \mu_i (t) \right) \, .
  \label{eq:current_density_with_dipoles}
  \end{equation}
In our case the oxygen, hydrogen, fluoride and chloride atoms carry the charges $-2e$, $e$, $-e$ and $-e$, respectively. 
In order to account for electronic polarization we conduct Wannier center localization.
Each Wannier center has the charge $-2e$. Four Wannier centers are unambiguously assigned to the closest heavy atom (O,F,Cl) and zero to hydrogen in agreement with the oktet rule.
Therefore, the electric dipole moment of atom $i$ is given by 
\begin{align}
  \mu_i(t) = -2 e \sum\limits_{j=1}^{N_i^\mathrm{WC}} \left( r^\mathrm{WC}_{ij}(t) - r_i(t) \right) \, .
  \label{eq:dipole_moment}
\end{align}
Here, $r^\mathrm{WC}_{ij}(t)$ is the position of the $j$-th Wannier center assigned to atom $i$, $r_i(t)$ is the position of atom $i$ and $N_i^\mathrm{WC}$ is the number of Wannier centers assigned to atom $i$. 
The distance vector $r^\mathrm{WC}_{ij}(t) - r_i(t)$ is defined as the shortest vector between the Wannier center and the atom, accounting for periodic boundary conditions.
We define the polarization density as
\begin{equation}
  P(t) = \frac{1}{V} \sum_{i=1}^{N} \left( q_i r_i(t) + \mu_i(t) \right) \, ,
\end{equation}
where $r_i(t)$ denotes the position of atom $i$. In contrast to $J(t)$ defined in Eq. \eqref{eq:current_density_with_dipoles}, $P(t)$ does depend on the choice of the atomic origins, which we choose to be the nuclei positions.
For consistency with the definition of the current density in Eq. \eqref{eq:current_density_with_dipoles},
the time derivative of the position must satisfy
$\dot{r}_i(t) = v_i(t)$. 
Under periodic boundary conditions, this requires the use of unwrapped atomic positions, i.e., positions that are made continuous in time by adding or subtracting the box length whenever an atom crosses a periodic boundary. 
We pick the initial atomic position $r_i(0)$ such that all atoms are inside the simulation box at time $t=0$, but other choices are possible as long as $r_i(t)$ is continuous in time.
In bulk systems, the external field in the perturbation Hamiltonian can be identified as the electric field \cite{sternCalculationDielectricPermittivity2003} and we can use the fluctuation dissipation theorem \cite{kuboFluctuationdissipationTheorem1966,carlsonExploringAbsorptionSpectrum2020} to relate the imaginary part of the dielectric function to the Fourier transformed trajectory of the polarization density of length $\tau$ as
\begin{align}
  \omega \tilde \varepsilon''(\omega) &= \omega^2 \frac{V}{2 \varepsilon_0 k_B T \tau }|\tilde{P}(\omega)|^2 \, .
  \label{eq:epsilon_imag_from_P}
\end{align} 
If the system is isotropic, Eq. \eqref{eq:epsilon_imag_from_P} can be averaged over the three dimensions.
Using Eq. \eqref{eq:current_density_polarization} we could also express the imaginary part of the dielectric function in terms of the current density as
\begin{align}
  \omega \tilde \varepsilon''(\omega) &= \frac{V}{2 \varepsilon_0 k_B T \tau }|\tilde{J}(\omega)|^2 \, .
  \label{eq:epsilon_imag_from_J}
\end{align}
Since the Wannier center localization procedure is numerically expensive, it is not numerically feasable to compute the atomic dipole moments at every time step. 
Therefore, $J(t)$ can only be computed from a highly discretized time series of $P(t)$, which leads to 
significant time discretization errors in Eq. \eqref{eq:epsilon_imag_from_J}.
This problem can be circumvented by using Eq. \eqref{eq:epsilon_imag_from_P} instead of Eq. \eqref{eq:epsilon_imag_from_J} to compute the imaginary part of the dielectric function, where time discretization is not an issue as long as the Nyquist–Shannon sampling criterion is fulfilled, i.e., as long as the sampling frequency exceeds twice the highest frequency present in the signal \cite{shannonCommunicationPresenceNoise1949}.
 In order to compute the absorption coefficient defined in Eq. \eqref{eq:absorption_coefficient} we also need the real part of the dielectric function, which is obtained by numerical Kramers-Kronig transformation as described elsewhere \cite{lehmannMultipolarElectricMagnetic2026} and where we add the high-frequency contribution $\varepsilon^\mathrm{VIS}=1.78$ \cite{haynesCRCHandbookChemistry2015a} to the real part of the dielectric function to account for electronic polarizability.
 \section{Extraction of the Difference Spectra from Experimental Data}
 \begin{figure}
  \centering
  \includegraphics[width=0.8\textwidth]{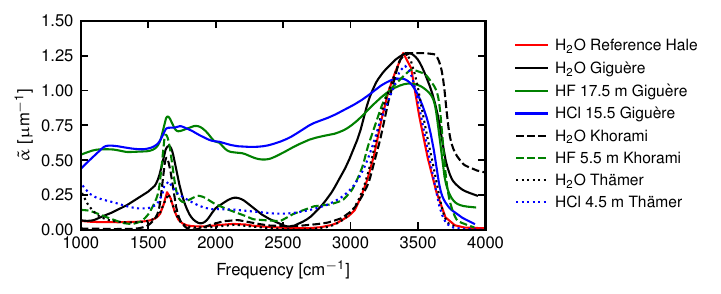}
  \caption{Comparison of experimental absorption spectra from the literature \cite{giguereH3OIonsAqueous1976,khoramiEtudeMelangesEau1987,thamerUltrafast2DIR2015}. Pure water spectra are shown in black, HF solutions in green, and HCl solutions in blue. For reference, the well-established literature absorption coefficient of pure water \cite{haleOpticalConstantsWater1973b} is shown in red and agrees best with the pure-water spectrum from Thämer \textit{et al.} \cite{thamerUltrafast2DIR2015}.
   While the experimental pure-water spectra differ considerably between the datasets, the spectral signatures of the excess protons are sufficiently pronounced and robust to enable a meaningful comparison across the different measurements.}  
  \label{fig:absorption_coefficient}
\end{figure}
All experimental spectra we compare our simulated spectra with are reported without units in the original works \cite{giguereH3OIonsAqueous1976,khoramiEtudeMelangesEau1987,thamerUltrafast2DIR2015}. In the dataset by Thämer et al. \cite{thamerUltrafast2DIR2015}, the zero baseline is clearly identifiable for the pure-water and the 4.5 m HCl spectra, therefore both spectra can be straightforwardly brought onto an absolute scale by multiplication with a factor that is chosen such that the maximum of the pure-water absorption spectrum for the OH-stretch peak matches the spectrum from Hale et al. \cite{haleOpticalConstantsWater1973b} for which units are given. As evident from Fig. S3, the pure water spectrum from Thämer et al. \cite{thamerUltrafast2DIR2015}, (black dotted line) agrees quantitatively with the spectrum from Ref. \cite{haleOpticalConstantsWater1973b} (red solid line), and therefore we consider it to be the most reliable dataset available. Except for the pure-water spectrum reported by Giguère and Turrell \cite{giguereH3OIonsAqueous1976}, for which we assume the boundary of the plotted spectral window to correspond to zero intensity, the spectra in Refs. ~\cite{giguereH3OIonsAqueous1976,khoramiEtudeMelangesEau1987} are vertically shifted by unknown offsets, and neither units nor zero lines are specified in the original works. Because pure water does not exhibit strong broad continuum bands (red line in Fig. S3), the pure-water spectrum of Khorami et al. \cite{khoramiEtudeMelangesEau1987} (black dashed line) is shifted such that its minimal intensity becomes equal to zero. In contrast, the broad continuum absorption of the acid spectra from Khorami et al. and Giguère and Turrell prevents the identification of a reliable minimum that could serve as a zero-intensity reference. Alignment at 4000~cm$^{-1}$ is likewise not justified because of the pronounced background, which has been attributed to reflectivity and interference effects in similar experimental setups \cite{ayotteWhyHydrofluoricAcid2005}. This background was also explicitly noted by Giguère and Turrell \cite{giguereNatureHydrofluoricAcid1980} and is evident from the substantial nonzero intensities of the
pure-water spectra near 4000 cm$^{-1}$ reported by Khorami et al. (black dashed line) and Giguère and Turrell (black solid line). In
the absence of a clearly identifiable absolute zero line for these acid-solution spectra, our procedure is to align them
based on the high-quality and zero-aligned 4.5 m HCl spectrum of Thämer et al. \cite{thamerUltrafast2DIR2015}, which is shown in Fig. S3 as a dotted blue line. In particular, we observe that HCl and HF solutions at comparable concentrations exhibit very similar line shapes. This holds for the 4.5 m HCl spectrum of Thämer et al. (dotted blue line) and the 5.5 m HF spectrum of Khorami et al. (dashed
green line), and is also apparent for the high-concentration HF and HCl spectra of Giguère and Turrell (solid green and
blue lines). We therefore align the 5.5~m HF spectrum of Khorami et al. with the 4.5~m HCl spectrum of Thämer et al., while the high-concentration HF and HCl spectra of Giguère and Turrell are aligned relative to each other. The remaining absolute position of
the zero line is chosen such that the overall continuum intensity scales approximately with acid concentration, using the Thämer
HCl data as the reference. The constant vertical offset is the only parameter required, since in the datasets of Khorami et al.
and Giguère and Turrell all spectra are shown in the same figure and therefore share the same intensity scale. The absolute
intensity scale of each dataset is determined unambiguously by multiplication with a single factor such that the maximum of the
corresponding pure-water spectrum matches the maximum of the Hale reference spectrum.
In fact, the uncertainty produced by our rescaling procedure could only be resolved by new experimental spectra that report a clearly identifiable absolute zero. Given the available data, our alignment procedure described above provides the most straightforward way of fixing this ambiguity. A residual constant offset between the HF and HCl spectra would amount to a frequency-independent contribution to the absorption, which would be difficult to justify physically. Also, our shifting of the zero line cannot be the cause of the observed agreement of the line shapes of HF and HCl. In particular, a constant vertical shift does not modify the relative intensities of the broad bands between 1000–1500 cm$^{-1}$, 1800–3000 cm$^{-1}$ and water OH-stretching band between 3000–4000 cm$^{-1}$, which are especially evident in the high-concentration HF and HCl spectra of Giguère and Turrell (solid green and blue lines in Fig. S3),
but can also be seen in the 5.5 m HF spectrum of Khorami et al. (dashed green line) and the 4.5 m HCl spectrum of Thämer et
al. (dotted blue line). We compute the experimental difference spectra by subtracting the corresponding pure-water spectrum
from each acid-solution spectrum according to Eq. (3) of the main text, using the experimental solution densities reported in
Refs. \cite{greenPerrysChemicalEngineers2019,haynesCRCHandbookChemistry2015a} for the normalization. Overall, although the quality of the absolute absorption spectra varies between the available
experimental datasets, the pronounced spectral signatures of the acids allow for robust comparison between our simulations and
the experimental data. Given the larger experimental uncertainties in the older datasets, future measurements of hydrohalic acids
at the same level as those from Thämer et al. \cite{thamerUltrafast2DIR2015} would be desirable to enable a direct quantitative analysis of the experimental spectral signatures of the excess proton in weak and strong acid solutions.
\color{black}
\bibliography{mybib}
\begin{table}[htbp]
  \centering
  \caption{Optimized Cartesian coordinates of the BLYP-D3 $\mathrm{H_2O}$ structure.}
  \label{tab:BLYP-water}
  \begin{tabular}{c S[table-format=-1.14] S[table-format=-1.14] S[table-format=-1.14]}
	\toprule
	atom & {$x$ [\AA]} & {$y$ [\AA]} & {$z$ [\AA]} \\
	\midrule
	O &  0.00000000000000 & -0.00000000004866 &  0.01146981887509 \\
	H &  0.00000000000000 &  0.76264088464080 & -0.59281490945719 \\
	H &  0.00000000000000 & -0.76264088459216 & -0.59281490941789 \\
	\bottomrule
\end{tabular}

\end{table}
\FloatBarrier

\begin{table}[htbp]
  \centering
  \caption{Optimized Cartesian coordinates of the BLYP-D3 $\mathrm{H_5O_2^+}$ structure.}
  \label{tab:BLYP-zundel}
  \begin{tabular}{c S[table-format=-1.14] S[table-format=-1.14] S[table-format=-1.14]}
	\toprule
	atom & {$x$ [\AA]} & {$y$ [\AA]} & {$z$ [\AA]} \\
	\midrule
	H &  0.19894201281841 & -0.00122473159411 & -0.00016454409742 \\
	O &  0.24486332136240 &  0.14788292318663 &  1.20955855775296 \\
	H &  0.19138320776292 & -0.72137611408892 &  1.65761770889904 \\
	H & -0.53070767925022 &  0.66324178132977 &  1.51299384550287 \\
	O &  0.24047634449361 & -0.15124660352796 & -1.21011718227015 \\
	H & -0.54287314082397 & -0.65585836892534 & -1.51163806471759 \\
	H &  0.19791593363686 &  0.71858111361993 & -1.65825032106971 \\
	\bottomrule
\end{tabular}

\end{table}
\FloatBarrier

\begin{table}[htbp]
  \centering
  \caption{Optimized Cartesian coordinates of the BLYP-D3 $\mathrm{H_3OF}$ structure.}
  \label{tab:BLYP-zundel-f}
  \begin{tabular}{c S[table-format=-1.14] S[table-format=-1.14] S[table-format=-1.14]}
	\toprule
	atom & {$x$ [\AA]} & {$y$ [\AA]} & {$z$ [\AA]} \\
	\midrule
	H & -0.16698537752274 & -0.06711170684854 & -0.15562333449468 \\
	O & -0.21429471267584 & -0.33235802377612 &  0.78182735016114 \\
	H &  0.81708068113579 &  0.44218280807118 &  1.58320219182464 \\
	H & -1.11034596166335 & -0.07417118639960 &  1.06761906849840 \\
	F &  1.50166737072612 &  0.92970110895307 &  2.11340072401048 \\
	\bottomrule
\end{tabular}

\end{table}
\FloatBarrier

\begin{table}[htbp]
  \centering
  \caption{Optimized Cartesian coordinates of the BLYP-D3 $\mathrm{H_3OCl}$ structure.}
  \label{tab:BLYP-zundel-cl}
  \begin{tabular}{c S[table-format=-1.14] S[table-format=-1.14] S[table-format=-1.14]}
	\toprule
	atom & {$x$ [\AA]} & {$y$ [\AA]} & {$z$ [\AA]} \\
	\midrule
	H & -0.15058428880015 & -0.15596825344804 & -0.14536122397240 \\
	O & -0.18802437863613 & -0.38273347900334 &  0.80395258017678 \\
	H &  0.73355263428805 &  0.55516354538902 &  1.51986673303624 \\
	H & -1.09842396294623 & -0.16006307491149 &  1.07867346927157 \\
	Cl &  1.61681324013949 &  1.40220622836145 &  2.20031311470466 \\
	\bottomrule
\end{tabular}

\end{table}
\FloatBarrier

\begin{table}[htbp]
  \centering
  \caption{Optimized Cartesian coordinates of the BLYP-D3 $\mathrm{H_9O_4^+}$ structure.}
  \label{tab:BLYP-eigen}
  \begin{tabular}{c S[table-format=-1.14] S[table-format=-1.14] S[table-format=-1.14]}
	\toprule
	atom & {$x$ [\AA]} & {$y$ [\AA]} & {$z$ [\AA]} \\
	\midrule
	O & -0.11016328636506 &  0.22553964214394 & -0.10070617731648 \\
	H & -0.02890898270197 &  0.26471599096013 &  0.93076086868733 \\
	H &  0.83692671550222 &  0.15186950311701 & -0.51176402298053 \\
	H & -0.65388856782313 & -0.61865714409080 & -0.35520042295749 \\
	O &  2.26419086172800 &  0.09748476743609 & -1.11323973268060 \\
	H &  2.56557586673845 & -0.82916285724742 & -1.15517106486515 \\
	H &  2.24296233421016 &  0.39355553018920 & -2.04238767154916 \\
	O & -1.46988478254727 & -1.86324058180704 & -0.77150119892174 \\
	H & -1.17678544745621 & -2.63063476938002 & -0.24538407690810 \\
	H & -2.40810333082126 & -1.74320660712673 & -0.53316136423892 \\
	O &  0.00934363677546 &  0.32065715786544 &  2.47783736643762 \\
	H &  0.70544552233430 & -0.27648687849232 &  2.80987757747302 \\
	H &  0.28242546042632 &  1.20903924643257 &  2.77341191982021 \\
	\bottomrule
\end{tabular}

\end{table}
\FloatBarrier

\begin{table}[htbp]
  \centering
  \caption{Optimized Cartesian coordinates of the BLYP-D3 $\mathrm{H_7O_3F}$ structure.}
  \label{tab:BLYP-eigen-f}
  \begin{tabular}{c S[table-format=-1.14] S[table-format=-1.14] S[table-format=-1.14]}
	\toprule
	atom & {$x$ [\AA]} & {$y$ [\AA]} & {$z$ [\AA]} \\
	\midrule
	O & -0.28375616749559 &  0.57499095868839 & -0.16220802421360 \\
	H & -0.12845383687477 &  0.41373614353071 &  0.80950180614806 \\
	H &  0.60645609325324 &  0.45092467092913 & -0.59360341853207 \\
	H & -1.17255537388874 & -0.40541691401180 & -0.65061044679161 \\
	O &  2.17952843284569 &  0.20816740836234 & -1.32874209563357 \\
	H &  2.82660592596424 &  0.25657183998710 & -0.60131276502036 \\
	H &  2.21078514940870 & -0.72456937053142 & -1.60949466866001 \\
	F & -1.83787713017201 & -1.10379009972976 & -1.01391750057798 \\
	O &  0.12327044726516 &  0.08460178098714 &  2.50778148814577 \\
	H &  0.72174685550543 & -0.68362158405356 &  2.54651227974475 \\
	H &  0.64844043710861 &  0.81309667298287 &  2.88613230974441 \\
	\bottomrule
\end{tabular}

\end{table}
\FloatBarrier

\begin{table}[htbp]
  \centering
  \caption{Optimized Cartesian coordinates of the BLYP-D3 $\mathrm{H_7O_3Cl}$ structure.}
  \label{tab:BLYP-eigen-cl}
  \begin{tabular}{c S[table-format=-1.14] S[table-format=-1.14] S[table-format=-1.14]}
	\toprule
	atom & {$x$ [\AA]} & {$y$ [\AA]} & {$z$ [\AA]} \\
	\midrule
	O & -0.04462702669227 & -0.82403921910089 & -0.23769821943493 \\
	H & -0.08678450995158 & -0.50447356796814 &  0.75549812731864 \\
	H &  0.80893244155144 & -0.42076164780206 & -0.66095190528580 \\
	H & -0.87482633593019 & -0.44056136778886 & -0.72138983412212 \\
	O &  2.10136604192149 &  0.15443567949868 & -1.30956494631188 \\
	H &  2.17697602077747 &  1.10137149475882 & -1.08812489316198 \\
	H &  2.88382149036002 & -0.25897882927418 & -0.89931836225362 \\
	O & -2.13577793200721 &  0.10692802095681 & -1.45073561496415 \\
	H & -2.22844170334817 &  1.05858829004548 & -1.25775366981808 \\
	H & -2.93496605081794 & -0.30188454508721 & -1.06912938497808 \\
	Cl & -0.17643243586305 &  0.08557969176154 &  2.53600470301202 \\
	\bottomrule
\end{tabular}

\end{table}
\FloatBarrier

\begin{table}[htbp]
  \centering
  \caption{Optimized Cartesian coordinates of the BLYP-D3 $\mathrm{H_5O_2F_2^-}$ structure.}
  \label{tab:BLYP-eigen-2f}
  \begin{tabular}{c S[table-format=-1.14] S[table-format=-1.14] S[table-format=-1.14]}
	\toprule
	atom & {$x$ [\AA]} & {$y$ [\AA]} & {$z$ [\AA]} \\
	\midrule
	O & -0.27568150961649 &  0.52337643888668 & -0.13676866419335 \\
	H & -0.07278571108308 &  0.32297894305106 &  0.90604103488969 \\
	H &  0.57827553198654 &  0.37927124733372 & -0.61895421233843 \\
	H & -1.13230636386651 & -0.28208341176158 & -0.59083882773904 \\
	O &  2.15413483192055 &  0.08421920169562 & -1.44410832248918 \\
	H &  2.82593962424080 &  0.58300646163210 & -0.94482105981266 \\
	H &  2.36505723920045 & -0.84797072305673 & -1.25454465461380 \\
	F & -1.91289877194212 & -0.97221565395483 & -0.99879714763046 \\
	F &  0.17929201815982 &  0.08285195327397 &  2.20975755162724 \\
	\bottomrule
\end{tabular}

\end{table}
\FloatBarrier
\begin{table}[htbp]
  \centering
  \caption{Optimized Cartesian coordinates of the BLYP-D3 $\mathrm{H_5O_2Cl_2^-}$ structure.}
  \label{tab:blyp-eigen-2cl}
  \begin{tabular}{c S[table-format=-1.14] S[table-format=-1.14] S[table-format=-1.14]}
	\toprule
	atom & {$x$ [\AA]} & {$y$ [\AA]} & {$z$ [\AA]} \\
	\midrule
	O &  0.10856672597814 & -0.99599209955845 & -0.03387014377664 \\
	H &  0.00375697014817 & -0.52492265594977 &  0.88981761205608 \\
	H &  0.87661823994882 & -0.52245882578204 & -0.53810466280513 \\
	H & -0.76917590838516 & -0.85747597427307 & -0.57937042676245 \\
	O &  2.02914053528928 &  0.17125071671328 & -1.32762744331743 \\
	H &  2.03431487263502 &  1.12354679753037 & -1.11683052979657 \\
	H &  2.89255929988974 & -0.15753131494375 & -1.01525474750054 \\
	Cl & -2.33748938366970 & -0.64557779766437 & -1.60933798973347 \\
	Cl & -0.12833187873432 &  0.35667560082783 &  2.55555058403616 \\
	\bottomrule
\end{tabular}

\end{table}
\FloatBarrier

\begin{table}[htbp]
  \centering
  \caption{Optimized Cartesian coordinates of the DSD-PBEP86-D3 $\mathrm{H_2O}$ structure.}
  \label{tab:DSD-PBEP86-water}
  \begin{tabular}{c S S S}
	\toprule
	atom & {$x$ [\AA]} & {$y$ [\AA]} & {$z$ [\AA]} \\
	\midrule
	O &  0.00000000000000 & -0.00000008168994 &  0.00573342356623 \\
	H &  0.00000000000000 &  0.75439146249341 & -0.58994674494221 \\
	H &  0.00000000000000 & -0.75439138080347 & -0.58994667862402 \\
	\bottomrule
\end{tabular}

\end{table}
\FloatBarrier

\begin{table}[htbp]
  \centering
  \caption{Optimized Cartesian coordinates of the DSD-PBEP86-D3 $\mathrm{H_5O_2^+}$ structure.}
  \label{tab:DSD-PBEP86-zundel}
  \begin{tabular}{c S S S}
	\toprule
	atom & {$x$ [\AA]} & {$y$ [\AA]} & {$z$ [\AA]} \\
	\midrule
	H &  0.19094079800485 & -0.00122185487313 & -0.00011263281192 \\
	O &  0.23613617088488 &  0.14291297542468 &  1.19004938359622 \\
	H &  0.19456337545845 & -0.71204845812109 &  1.64267712289641 \\
	H & -0.52116898919560 &  0.65828441843909 &  1.50399302462895 \\
	O &  0.23179357960223 & -0.14614864804201 & -1.19057918666067 \\
	H & -0.53324946295829 & -0.65102515240032 & -1.50279346412270 \\
	H &  0.20098452820350 &  0.70924671957278 & -1.64323424752629 \\
	\bottomrule
\end{tabular}

\end{table}
\FloatBarrier

\begin{table}[htbp]
  \centering
  \caption{Optimized Cartesian coordinates of the DSD-PBEP86-D3 $\mathrm{H_3OF}$ structure.}
  \label{tab:DSD-PBEP86-zundel-f}
  \begin{tabular}{c S S S}
	\toprule
	atom & {$x$ [\AA]} & {$y$ [\AA]} & {$z$ [\AA]} \\
	\midrule
	H & -0.17649631971899 & -0.07722383121088 & -0.15156972511996 \\
	O & -0.21472135567348 & -0.31316287245075 &  0.78220211097013 \\
	H &  0.82858392313273 &  0.45417859739939 &  1.59149232188196 \\
	H & -1.10995565033560 & -0.08479302062344 &  1.05663810461041 \\
	F &  1.49971140259533 &  0.91924412688567 &  2.11166318765745 \\
	\bottomrule
\end{tabular}

\end{table}
\FloatBarrier

\begin{table}[htbp]
  \centering
  \caption{Optimized Cartesian coordinates of the DSD-PBEP86-D3 $\mathrm{H_3OCl}$ structure.}
  \label{tab:DSD-PBEP86-zundel-cl}
  \begin{tabular}{c S S S}
	\toprule
	atom & {$x$ [\AA]} & {$y$ [\AA]} & {$z$ [\AA]} \\
	\midrule
	H & -0.17237207805234 & -0.17785885220707 & -0.14943508776723 \\
	O & -0.19747013921092 & -0.36873666671446 &  0.79604331505396 \\
	H &  0.76426335795241 &  0.59140651269666 &  1.54179428296405 \\
	H & -1.10627100037363 & -0.18084037600534 &  1.06003972076303 \\
	Cl &  1.62518310372950 &  1.39463434861781 &  2.20900244220305 \\
	\bottomrule
\end{tabular}

\end{table}
\FloatBarrier

\begin{table}[htbp]
  \centering
  \caption{Optimized Cartesian coordinates of the DSD-PBEP86-D3 $\mathrm{H_9O_4^+}$ structure.}
  \label{tab:DSD-PBEP86-eigen}
  \begin{tabular}{c S S S}
	\toprule
	atom & {$x$ [\AA]} & {$y$ [\AA]} & {$z$ [\AA]} \\
	\midrule
O & -0.07564559192674 &  0.16357864403942 & -0.06848527595678 \\
H &  0.01719166643526 &  0.21205794330723 &  0.94249982649813 \\
H &  0.84267167616664 &  0.10435499318931 & -0.49865505493442 \\
H & -0.62672010262729 & -0.65337634006228 & -0.31884973340391 \\
O &  2.24562656901349 &  0.08934429930945 & -1.13551377148416 \\
H &  2.60152497296763 & -0.80328603636570 & -1.21159460305849 \\
H &  2.23997083860774 &  0.43099941497492 & -2.03687341148738 \\
O & -1.47105711110599 & -1.86021331698638 & -0.75284515682213 \\
H & -1.30907939045878 & -2.63332894446410 & -0.20024520432539 \\
H & -2.41679316232786 & -1.69171330938612 & -0.67185973020329 \\
O &  0.06149889526922 &  0.28698583939852 &  2.47785588326534 \\
H &  0.84574936928121 & -0.13966226426681 &  2.84126253501360 \\
H &  0.10419737070547 &  1.19573207731258 &  2.79667569689887 \\
\bottomrule
\end{tabular}

\end{table}
\FloatBarrier

\begin{table}[htbp]
  \centering
  \caption{Optimized Cartesian coordinates of the DSD-PBEP86-D3 $\mathrm{H_7O_3F}$ structure.}
  \label{tab:DSD-PBEP86-eigen-f}
  \begin{tabular}{c S S S}
	\toprule
	atom & {$x$ [\AA]} & {$y$ [\AA]} & {$z$ [\AA]} \\
	\midrule
O & -0.21586023755127 &  0.52429414546557 & -0.14378518485308 \\
H & -0.07061121544156 &  0.36028114582630 &  0.81110007672530 \\
H &  0.64379023330182 &  0.35153856476882 & -0.58082642837674 \\
H & -1.23749034085213 & -0.34221341248185 & -0.65759579349723 \\
O &  2.14708355390966 &  0.05387512638098 & -1.44198637210836 \\
H &  2.84547628462956 &  0.63409786319727 & -1.12137076333759 \\
H &  2.47185695254037 & -0.83245509331408 & -1.25209028445931 \\
F & -1.95867707442826 & -0.92242210364671 & -1.01931350784008 \\
O &  0.08345873302971 &  0.04643822082211 &  2.53283395545190 \\
H &  0.62178508614783 & -0.73817214646115 &  2.68019404763230 \\
H &  0.56337885763420 &  0.74942919658388 &  2.98287921901669 \\
\bottomrule
\end{tabular}

\end{table}
\FloatBarrier

\begin{table}[htbp]
  \centering
  \caption{Optimized Cartesian coordinates of the DSD-PBEP86-D3 $\mathrm{H_7O_3Cl}$ structure.}
  \label{tab:DSD-PBEP86-eigen-cl}
  \begin{tabular}{c S S S}
	\toprule
	atom & {$x$ [\AA]} & {$y$ [\AA]} & {$z$ [\AA]} \\
	\midrule
O & -0.04834522273317 & -0.78357571317450 & -0.21906284216951 \\
H & -0.08613254636449 & -0.49235724738344 &  0.75678673929421 \\
H &  0.79537327827391 & -0.39862878592522 & -0.63671932823418 \\
H & -0.86690146928360 & -0.41179760810877 & -0.69519148465888 \\
O &  2.08016665781174 &  0.14549095884476 & -1.29284152890735 \\
H &  2.19502553756192 &  1.08630100762655 & -1.11591078942679 \\
H &  2.87017796957679 & -0.27188332514189 & -0.93074423371143 \\
O & -2.11166527835309 &  0.11095041502356 & -1.43599463667036 \\
H & -2.25887226910972 &  1.04778611179663 & -1.26201198940433 \\
H & -2.91640922597618 & -0.32573217330111 & -1.13385490703482 \\
Cl & -0.16317743140409 &  0.04965035974342 &  2.56238100092347 \\
\bottomrule
\end{tabular}

\end{table}
\FloatBarrier

\begin{table}[htbp]
  \centering
  \caption{Optimized Cartesian coordinates of the DSD-PBEP86-D3 $\mathrm{H_5O_2F_2^-}$ structure.}
  \label{tab:DSD-PBEP86-eigen-2f}
  \begin{tabular}{c S S S}
	\toprule
	atom & {$x$ [\AA]} & {$y$ [\AA]} & {$z$ [\AA]} \\
	\midrule
O & -0.10092856551595 &  0.39749901492049 & -0.17551729135083 \\
H & -0.04532032342395 &  0.40509416213689 &  0.79748057910531 \\
H &  0.87127392844466 &  0.25619596079241 & -0.52505495805713 \\
H & -0.91187783882859 & -0.57821422978706 & -0.56998395082908 \\
F &  2.15391807542377 &  0.07672432068768 & -0.99905147482847 \\
H &  0.57867402271897 & -0.37154042418136 &  2.87936602251112 \\
H &  0.56811274089971 &  1.14494986761447 &  2.91874819853170 \\
F & -1.56075130504505 & -1.32408219734263 & -0.89298988731807 \\
O &  0.05482424822644 &  0.39057631405910 &  2.61140247733544 \\
\bottomrule
\end{tabular}

\end{table}
\FloatBarrier
\begin{table}[htbp]
  \centering
  \caption{Optimized Cartesian coordinates of the DSD-PBEP86-D3 $\mathrm{H_5O_2Cl_2^-}$ structure.}
  \label{tab:DSD-PBEP86-eigen-2cl}
  \begin{tabular}{c S S S}
	\toprule
	atom & {$x$ [\AA]} & {$y$ [\AA]} & {$z$ [\AA]} \\
	\midrule
O & 0.11996235616092 & -0.94186849765233 & -0.00408042402472 \\
H & -0.03069953003691 & -0.48249135338552 & 0.89337419248553 \\
H & 0.86692112465704 & -0.4550267017101 & -0.49428345344853 \\
H & -0.73311374944769 & -0.86534535081708 & -0.55731131677956 \\
O & 1.98447272464396 & 0.2486911619816 & -1.29003800410644 \\
H & 2.06730403097527 & 1.17674856103277 & -1.04232175632205 \\
H & 2.85300293641895 & -0.13226251165643 & -1.116784754918 \\
Cl & -2.2960113533832 & -0.75332665868051 & -1.61330501483688 \\
Cl & -0.27132553998834 & 0.3555503508876 & 2.57081353195065 \\
\bottomrule
\end{tabular}

\end{table}
\FloatBarrier

\begin{table}[htbp]
  \centering
  \caption{Optimized Cartesian coordinates of the MP2 $\mathrm{H_2O}$ structure.}
  \label{tab:MP2-water}
  \begin{tabular}{c S S S}
	\toprule
	atom & {$x$ [\AA]} & {$y$ [\AA]} & {$z$ [\AA]} \\
	\midrule
	O &  0.00000000000000 &  0.00000000066133 &  0.00765288921956 \\
	H &  0.00000000000000 &  0.75232481818908 & -0.59090644515089 \\
	H &  0.00000000000000 & -0.75232481885041 & -0.59090644406867 \\
	\bottomrule
\end{tabular}

\end{table}
\FloatBarrier

\begin{table}[htbp]
  \centering
  \caption{Optimized Cartesian coordinates of the MP2 $\mathrm{H_5O_2^+}$ structure.}
  \label{tab:MP2-zundel}
  \begin{tabular}{c S S S}
	\toprule
	atom & {$x$ [\AA]} & {$y$ [\AA]} & {$z$ [\AA]} \\
	\midrule
	H &  0.19354819599318 & -0.00112311946203 & -0.00006906258208 \\
	O &  0.23951092625067 &  0.14530018457204 &  1.18904161554620 \\
	H &  0.19320057105368 & -0.71149326253664 &  1.63956193419482 \\
	H & -0.52455012194912 &  0.65474884429824 &  1.49884144298160 \\
	O &  0.23517361087821 & -0.14853578450077 & -1.18954586193826 \\
	H & -0.53655637730350 & -0.64747667874477 & -1.49742097110756 \\
	H &  0.19967319507691 &  0.70857981637394 & -1.64040909709472 \\
	\bottomrule
\end{tabular}

\end{table}
\FloatBarrier

\begin{table}[htbp]
  \centering
  \caption{Optimized Cartesian coordinates of the MP2 $\mathrm{H_3OF}$ structure.}
  \label{tab:MP2-zundel-f}
  \begin{tabular}{c S S S}
	\toprule
	atom & {$x$ [\AA]} & {$y$ [\AA]} & {$z$ [\AA]} \\
	\midrule
	H & -0.17752664243435 & -0.07602096526288 & -0.15029101990065 \\
	O & -0.21283642114328 & -0.31487283142843 &  0.78348317807833 \\
	H &  0.82740251882024 &  0.45291933842279 &  1.59049723499250 \\
	H & -1.10861628468019 & -0.08331708398670 &  1.05560383662291 \\
	F &  1.49869882943758 &  0.91953454225520 &  2.11113277020691 \\
	\bottomrule
\end{tabular}

\end{table}
\FloatBarrier

\begin{table}[htbp]
  \centering
  \caption{Optimized Cartesian coordinates of the MP2 $\mathrm{H_3OCl}$ structure.}
  \label{tab:MP2-zundel-cl}
  \begin{tabular}{c S S S}
	\toprule
	atom & {$x$ [\AA]} & {$y$ [\AA]} & {$z$ [\AA]} \\
	\midrule
	H & -0.17403277840485 & -0.17752215158754 & -0.14868595747086 \\
	O & -0.19560394309150 & -0.36823377973086 &  0.79768449440409 \\
	H &  0.76586907552056 &  0.59237589700479 &  1.54353967850171 \\
	H & -1.10649861836079 & -0.18111091917883 &  1.05772968248353 \\
	Cl &  1.62359950838160 &  1.39309591988004 &  2.20717677529839 \\
	\bottomrule
\end{tabular}

\end{table}
\FloatBarrier

\begin{table}[htbp]
  \centering
  \caption{Optimized Cartesian coordinates of the MP2 $\mathrm{H_9O_4^+}$ structure.}
  \label{tab:MP2-eigen}
  \begin{tabular}{c S S S}
	\toprule
	atom & {$x$ [\AA]} & {$y$ [\AA]} & {$z$ [\AA]} \\
	\midrule
	O & -0.08004368527878 &  0.17117504012599 & -0.06725335165636 \\
	H &  0.01293368326955 &  0.21616177773996 &  0.94419806797308 \\
	H &  0.84013911435555 &  0.10598282590156 & -0.49344910816888 \\
	H & -0.62787576229505 & -0.64871430458717 & -0.31711821283322 \\
	O &  2.24103493237243 &  0.08422410045934 & -1.12986874783648 \\
	H &  2.58816753604723 & -0.81183692213440 & -1.21411709391472 \\
	H &  2.23293823280577 &  0.42493777228168 & -2.03230395637866 \\
	O & -1.46223928453256 & -1.85554727548494 & -0.75872049955539 \\
	H & -1.29288707668640 & -2.63311298135008 & -0.21346336046899 \\
	H & -2.40975369162971 & -1.69885423681994 & -0.66758770788546 \\
	O &  0.05777099573741 &  0.29180723295482 &  2.47754709337622 \\
	H &  0.83696354220568 & -0.14541800485657 &  2.84107393694357 \\
	H &  0.12198746362886 &  1.20066797576979 &  2.79443494040528 \\
	\bottomrule
\end{tabular}

\end{table}
\FloatBarrier

\begin{table}[htbp]
  \centering
  \caption{Optimized Cartesian coordinates of the MP2 $\mathrm{H_7O_3F}$ structure.}
  \label{tab:MP2-eigen-f}
  \begin{tabular}{c S S S}
	\toprule
	atom & {$x$ [\AA]} & {$y$ [\AA]} & {$z$ [\AA]} \\
	\midrule
	O & -0.22004669278772 &  0.52466615321551 & -0.14552750537934 \\
	H & -0.07197068374216 &  0.35910140124324 &  0.80896806818114 \\
	H &  0.64108909169890 &  0.35236060054872 & -0.58052460792064 \\
	H & -1.23635942214845 & -0.34022548829382 & -0.65787581183401 \\
	O &  2.14472477267237 &  0.05343505668062 & -1.44049947743632 \\
	H &  2.84536455733006 &  0.63205182034811 & -1.12020952675114 \\
	H &  2.47347397625657 & -0.83130677284743 & -1.24735198680113 \\
	F & -1.95784809974782 & -0.92240584818769 & -1.02051670320343 \\
	O &  0.08370851535277 &  0.04639967309040 &  2.53108859532349 \\
	H &  0.62514597506088 & -0.73646568339865 &  2.68008258485101 \\
	H &  0.56690884297454 &  0.74708059474213 &  2.98240533532417 \\
	\bottomrule
\end{tabular}

\end{table}
\FloatBarrier

\begin{table}[htbp]
  \centering
  \caption{Optimized Cartesian coordinates of the MP2 $\mathrm{H_7O_3Cl}$ structure.}
  \label{tab:MP2-eigen-cl}
  \begin{tabular}{c S S S}
	\toprule
	atom & {$x$ [\AA]} & {$y$ [\AA]} & {$z$ [\AA]} \\
	\midrule
	O & -0.04797826809181 & -0.79271798026414 & -0.22145380168680 \\
	H & -0.08630964036133 & -0.49353381490214 &  0.75317069814926 \\
	H &  0.79437097864812 & -0.40216247310967 & -0.63803221632450 \\
	H & -0.86532876025775 & -0.41588176459751 & -0.69671032381437 \\
	O &  2.07545662577765 &  0.14653349116190 & -1.29144950924611 \\
	H &  2.19089616745239 &  1.08666151772245 & -1.10773350615728 \\
	H &  2.86739895263589 & -0.26791106752368 & -0.92833854402311 \\
	O & -2.10768117609832 &  0.11058238506802 & -1.43435537449341 \\
	H & -2.25372408538585 &  1.04766784922317 & -1.25701983831393 \\
	H & -2.91436066587530 & -0.32055240468177 & -1.12730304595399 \\
	Cl & -0.16350012844367 &  0.05751826190336 &  2.54606146186425 \\
	\bottomrule
\end{tabular}

\end{table}
\FloatBarrier

\begin{table}[htbp]
  \centering
  \caption{Optimized Cartesian coordinates of the MP2 $\mathrm{H_5O_2F_2^-}$ structure.}
  \label{tab:MP2-eigen-2f}
  \begin{tabular}{c S S S}
	\toprule
	atom & {$x$ [\AA]} & {$y$ [\AA]} & {$z$ [\AA]} \\
	\midrule
	O & -0.10437962229873 &  0.40109969865654 & -0.18041895777505 \\
	H & -0.04736968496674 &  0.40543005333087 &  0.79316066248525 \\
	H &  0.87143991393022 &  0.25584029430888 & -0.52722797566355 \\
	H & -0.90326379290198 & -0.57056057216015 & -0.56388102811323 \\
	F &  2.14765966288366 &  0.07428288698048 & -0.99418662229154 \\
	H &  0.57641651920338 & -0.37281497185682 &  2.87669814837706 \\
	H &  0.56918345034624 &  1.14027576940850 &  2.91820168655413 \\
	F & -1.55485722433826 & -1.32560709982388 & -0.88371274480209 \\
	O &  0.05309576104222 &  0.38925673005556 &  2.60576654632902 \\
	\bottomrule
\end{tabular}

\end{table}
\FloatBarrier
\begin{table}[htbp]
  \centering
  \caption{Optimized Cartesian coordinates of the MP2 $\mathrm{H_5O_2Cl_2^-}$ structure.}
  \label{tab:MP2-eigen-2cl}
  \begin{tabular}{c S S S}
	\toprule
	atom & {$x$ [\AA]} & {$y$ [\AA]} & {$z$ [\AA]} \\
	\midrule
	O &  0.12267038099215 & -0.95299569712591 & -0.00355392307107 \\
	H & -0.02869063713668 & -0.48506736382637 &  0.89093989703971 \\
	H &  0.86713953456617 & -0.46121220629685 & -0.49422150159096 \\
	H & -0.73099224379019 & -0.86738227157759 & -0.55701465099311 \\
	O &  1.97815635360751 &  0.24797611784173 & -1.28890203215115 \\
	H &  2.05929143301248 &  1.17552698186463 & -1.03597605524349 \\
	H &  2.84974011687314 & -0.12663413665811 & -1.11321168884914 \\
	Cl & -2.28687519023699 & -0.74122777071567 & -1.60410248456860 \\
	Cl & -0.26992674788759 &  0.36168534649414 &  2.55210543942781 \\
	\bottomrule
\end{tabular}

\end{table}
\FloatBarrier

\begin{table}[htbp]
  \centering
  \caption{Optimized Cartesian coordinates of the CCSD(T) $\mathrm{H_2O}$ structure.}
  \label{tab:CCSD(T)-water}
  \begin{tabular}{c S S S}
	\toprule
	atom & {$x$ [\AA]} & {$y$ [\AA]} & {$z$ [\AA]} \\
	\midrule
O &  0.00000000000000 &  0.00000000388830 &  0.00723391153094 \\
H &  0.00000000000000 &  0.75277566112357 & -0.59069695508998 \\
H &  0.00000000000000 & -0.75277566501187 & -0.59069695644096 \\
\bottomrule
\end{tabular}

\end{table}
\FloatBarrier

\begin{table}[htbp]
  \centering
  \caption{Optimized Cartesian coordinates of the CCSD(T) $\mathrm{H_5O_2^+}$ structure.}
  \label{tab:CCSD(T)-zundel}
  \begin{tabular}{c S S S}
	\toprule
	atom & {$x$ [\AA]} & {$y$ [\AA]} & {$z$ [\AA]} \\
	\midrule
	H &  0.19312696926305 & -0.00125098699713 & -0.00008661376854 \\
	O &  0.23858423138102 &  0.14520865302191 &  1.18794314928745 \\
	H &  0.19488406075107 & -0.71078426476448 &  1.63884221253557 \\
	H & -0.52505988115399 &  0.65400017919093 &  1.49783543852249 \\
	O &  0.23422322521033 & -0.14845208911950 & -1.18848462163853 \\
	H & -0.53706331971874 & -0.64669068299670 & -1.49658009212427 \\
	H &  0.20130471426728 &  0.70796919166499 & -1.63946947281417 \\
	\bottomrule
\end{tabular}

\end{table}
\FloatBarrier

\begin{table}[htbp]
  \centering
  \caption{Optimized Cartesian coordinates of the CCSD(T) $\mathrm{H_3OF}$ structure.}
  \label{tab:CCSD(T)-zundel-f}
  \begin{tabular}{c S S S}
	\toprule
	atom & {$x$ [\AA]} & {$y$ [\AA]} & {$z$ [\AA]} \\
	\midrule
	H & -0.17962120723381 & -0.07648213708241 & -0.15180009680941 \\
	O & -0.21642239184846 & -0.31850100355603 &  0.78061570655865 \\
	H &  0.83310115983558 &  0.45631493828911 &  1.59502985583042 \\
	H & -1.11045428848275 & -0.08369651882284 &  1.05404099405226 \\
	F &  1.50051872772943 &  0.92060772117217 &  2.11253954036807 \\
	\bottomrule
\end{tabular}

\end{table}
\FloatBarrier

\begin{table}[htbp]
  \centering
  \caption{Optimized Cartesian coordinates of the CCSD(T) $\mathrm{H_3OCl}$ structure.}
  \label{tab:CCSD(T)-zundel-cl}
  \begin{tabular}{c S S S}
	\toprule
	atom & {$x$ [\AA]} & {$y$ [\AA]} & {$z$ [\AA]} \\
	\midrule
	H & -0.19501502464966 & -0.19153062694403 & -0.16292438645882 \\
	O & -0.21661015972887 & -0.38675891421099 &  0.78126300851517 \\
	H &  0.80113412666915 &  0.62202928674708 &  1.57123300857690 \\
	H & -1.12531603807257 & -0.19499898146180 &  1.04104385754966 \\
	Cl &  1.64914033982696 &  1.40986420225734 &  2.22682918503396 \\
	\bottomrule
\end{tabular}

\end{table}
\FloatBarrier